\documentclass[aps,pra,showpacs,twocolumn]{revtex4-2}
\usepackage{bm}
\usepackage{amsmath}
\usepackage{amssymb}
\usepackage{slashed}
\usepackage{float}
\allowdisplaybreaks
\usepackage{subcaption}
\usepackage{dcolumn}
\usepackage{graphicx}
\newcolumntype{.}{D{x}{}{-1}}
\newcommand*{\cent}[1]{\multicolumn{1}{c}{$#1$}}

\newcolumntype{w}[1]{D{.}{.}{#1}}
\newcolumntype{L}{>{$}l<{$}}
\newcommand{\balpha}{\vec{\alpha}}
\newcommand{\bgamma}{\vec{\gamma}}
\newcommand{\bsigma}{\vec{\sigma}}
\newcommand{\bfx}{\vec{x}}
\newcommand{\bfy}{\vec{y}}
\newcommand{\bfp}{\vec{p}}
\newcommand{\bfq}{\vec{q}}
\newcommand{\bfr}{\vec{r}}

\newcommand{\bfz}{\vec{z}}
\newcommand{\Za}{{Z\alpha}}

\usepackage{graphicx}
\usepackage{xcolor}
\usepackage{nicefrac}
\usepackage{physics}
\usepackage{multirow}

\newcommand{\lbr}{\langle}
\newcommand{\rbr}{\rangle}

\newcommand{\vare}{\varepsilon}
\newcommand{\crossed}[1]{#1\!\!\!/}

\newcommand{\hp}{\hat{\vec{p}}}
\newcommand{\hq}{\hat{\vec{q}}}

\newcommand{\rp}{{\rm p}}

\begin{document}

\title{Two-loop electron self-energy in bound-electron $\bm{g}$ factor:
diagrams in momentum-coordinate representation}

\author{V.~A. Yerokhin}
\email{Corresponding author: vladimir.yerokhin@mpi-hd.mpg.de}
\affiliation{Max~Planck~Institute for Nuclear Physics, Saupfercheckweg~1, D~69117 Heidelberg, Germany}

\author{B.~Sikora}
\affiliation{Max~Planck~Institute for Nuclear Physics, Saupfercheckweg~1, D~69117 Heidelberg, Germany}

\author{Z. Harman}
\affiliation{Max~Planck~Institute for Nuclear Physics, Saupfercheckweg~1, D~69117 Heidelberg, Germany}

\author{C.~H. Keitel}
\affiliation{Max~Planck~Institute for Nuclear Physics, Saupfercheckweg~1, D~69117 Heidelberg, Germany}

\begin{abstract}

The two-loop electron self-energy correction is one of the most problematic QED effects and,
for a long time, was the dominant source of uncertainty in the theoretical prediction
of the bound-electron $g$ factor in hydrogen-like ions.
A major breakthrough was recently achieved in [B.~Sikora et al. Phys. Rev. Lett. 134, 123001 (2025)],
where this effect was calculated without any expansion in the nuclear binding strength
parameter $\Za$ (where $Z$ is the nuclear charge number and $\alpha$ is the fine-structure
constant). In this paper, we describe our calculations of one of the most difficult parts
of the two-loop self-energy, represented by Feynman diagrams that are treated in the
mixed momentum-coordinate representation.

\end{abstract}

\maketitle

\section{Introduction}

The $g$ factor of the bound electron in hydrogen-like ions has been measured with
exceptional accuracy of up to 11 significant figures
\cite{haeffner:00:prl,verdu:04,sturm:11,sturm:13:Si,sturm:14,sailer:22,morgner:23},
thus challenging theorists to match this precision in the bound-state QED
calculations. After many-years efforts and dedicated calculations,
theory has been able to reach the $10^{-11}$ accuracy level for light
hydrogen-like ions, such as carbon and oxygen
\cite{persson:97:g,beier:00:rep,shabaev:02:recprl,yerokhin:02:prl,pachucki:04:prl,pachucki:05:gfact}.
For higher-$Z$ ions, however, theory suffered from a dramatic loss
of precision, mostly from uncalculated two-loop QED contributions.
There have been significant efforts invested
in calculations of two-loop QED effects, both within expansion in the parameter
$\Za$ \cite{czarnecki:16,czarnecki:18,czarnecki:20}
and without \cite{yerokhin:13:twoloopg,debierre:21} in last years.
They resulted in a better theoretical understanding of the $g$ factors
in the low-$Z$ region
but did not significantly improved theoretical accuracy
in the high-$Z$ region. In particular,
for the heaviest measured element, tin ($Z = 50$), a theoretical
accuracy of about $10^{-7}$
was recently reported in Ref.~\cite{morgner:23}, substantially inferior to
the $10^{-10}$ experimental accuracy achieved in the same
work.
The main factor limiting the theoretical accuracy was the
two-loop self-energy correction.
Because its $\Za$ expansion converges slowly, any substantial progress for
the $g$ factor of tin (as well as heavier elements)
could be achieved only through a calculation nonperturbative in the parameter  $\Za$.

The two-loop self-energy correction has been calculated without expansion
in $\Za$ for the {\em Lamb shift} in a series of investigations
\cite{yerokhin:03:prl,yerokhin:05:sese,yerokhin:06:prl,yerokhin:09:sese,yerokhin:18:sese,yerokhin:24:sese,yerokhin:25:sese}.
Carrying out analogous calculations for the bound-electron $g$ factor is even more challenging. 
The corresponding project was initiated about a decade ago.
Several individual contributions have been computed \cite{sikora:18:phd,sikora:20}
but two of the most difficult parts of the two-loop self-energy correction,
the so-called $M$ and $P$ terms, remained uncalculated.
A breakthrough was achieved in our recent Letter
\cite{sikora:25}, where the complete calculation
of the two-loop self-energy correction to the $g$-factor was reported for
the hydrogen-like tin
ion. This calculation enhanced theoretical accuracy by
an order of magnitude and significantly improved agreement of the
theoretical prediction of the tin $g$ factor
with the experimental value \cite{morgner:23}.

The goal of the present investigation is to present our method developed
for the $P$-term part of the two-loop self-energy correction and perform
numerical calculations for hydrogen-like ions with $Z \ge 50$. The corresponding
Feynman diagrams contain both the Dirac-Coulomb propagators and
ultraviolet-diverging free-electron subgraphs. For this reason,
their calculation needs to be performed in the mixed momentum-coordinate
representation. Such contributions are characteristic for non-factorizable
two-loop QED effects and were encountered for the first time
in the Lamb-shift calculations of the two-loop self-energy.
In early studies
\cite{yerokhin:01:sese,yerokhin:03:epjd}, they were computed by using
the $B$-spline finite-basis representation of the Dirac-Coulomb spectrum and the
subsequent numerical Fourier transformation. Later on, a different technique
was developed \cite{yerokhin:10:sese} which used the
analytical representation of the Dirac-Coulomb Green function
in terms of Whittaker functions. This technique allowed to
improve the numerical precision and obtain
accurate results for the two-loop Lamb shift \cite{yerokhin:24:sese,yerokhin:25:sese}.
In the present work, we extend the technique developed for the Lamb shift
to the case of the bound-electron $g$ factor.

The relativistic units ($\hbar=c=m=1$) and the Heaviside charge units ($ \alpha = e^2/4\pi$, $e<0$)
are used throughout this paper.
We use roman style ($\rp$) for four vectors,
arrows $\bfp$ for three
vectors, and italic style ($p$) for scalars.
Four vectors have the form $\rp = (\rp_0,\bfp)$.
The Feynman gauge will be used in this work.

\section{Basic formulation}

The $g$ factor of an atomic system is observed as a shift of atomic energy levels
due to the interaction with a weak external magnetic field (the linear Zeeman effect).
The interaction potential is
\begin{align}
V_{\rm Zee}(\bfr) = -e\,\balpha\cdot\vec{A}(\bfr)\,,
\end{align}
where $\vec{A}(\bfr) = (1/2)[\vec{B}\times\bfr]$ is the vector potential,
$\balpha$ is the vector of Dirac matrices and $e$ is the electron charge.
In practical calculations, it is convenient to introduce the
effective $g$-factor potential $V_{g}$ \cite{yerokhin:10:sehfs},
\begin{align}
\label{eq:00}
V_{g}(\bfr) = \frac1{\mu_a}\,(\bfr \times \balpha)_z\,,
\end{align}
where $\mu_a$ denotes the angular momentum projection of the reference state and the
subscript $z$ indicates the projection on the $z$ axis.
The effective potential $V_g$ does not depend on the magnetic field and its expectation value
yields directly the $g$-factor value. So, for the $1s$ reference state and
point nucleus, one obtains the well-known Dirac value of a bound-electron $g$ factor,
\begin{align}
\label{eq:01}
g_{\rm D} =  \bra{a} V_g \ket{a} = \frac23\, \Big( 1 + 2\sqrt{1-(\Za)^2}\Big)\,.
\end{align}

\begin{figure}
\centerline{
\resizebox{\columnwidth}{!}{%
  \includegraphics{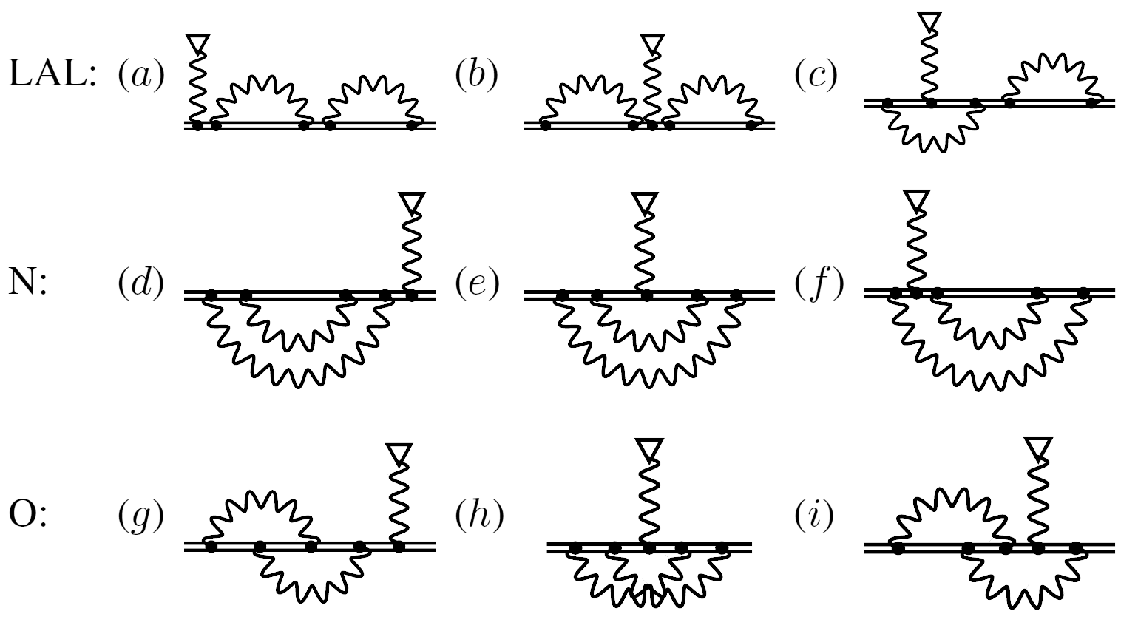}
}}
 \caption{
Feynman diagrams representing the two-loop electron self-energy (SESE)
correction
for the bound-electron $g$ factor:
({\em a})-({\em c}) loop-after-loop (LAL), ({\em d})-({\em f}) nested (N), ({\em g})-({\em i}) overlapping (O)
diagrams.
The double line denotes the electron in the
presence of the binding nuclear field; the wavy line indicates the exchange of
a virtual photon; the wavy line terminated by a triangle denotes
the interaction with the external magnetic field.
\label{fig:sese}}
\end{figure}

The two-loop self-energy (SESE) correction to the $g$ factor of a bound electron is
graphically represented in Fig.~\ref{fig:sese}. Feynman diagrams in the first
line are known as the loop-after-loop (LAL) diagrams. They can be factorized in
terms of the self-energy contributions of lower orders. Their analysis and
calculation was reported in our previous investigations
\cite{sikora:18:phd,sikora:20}. In the present work we will be concerned
with the nested and overlapping diagrams shown in the second and the third lines
of Fig.~\ref{fig:sese}, respectively.

The nested (N) and overlapping (O) diagrams in Fig.~\ref{fig:sese} are of two types:
(i)
with the magnetic interaction attached to the external wave function
[diagrams ($d$) and ($g$)] and (ii)
with the magnetic interaction attached
to the electron propagators [diagrams ($e$), ($f$), ($h$) and ($i$)].
Contributions of diagrams of first type
are divided into the irreducible (with $n\neq a$) and the reducible
(with $n = a$)
parts, where $n$ denotes an intermediate electron state in the propagator next
to the magnetic interaction and $a$ is the reference state. The irreducible contributions
will be referred to as the {\em perturbed-orbital} (W) terms. They have
the following structure
\begin{align}
\label{eq:02}
E_{h{\rm W}} = \bra{a} \Sigma_{h} \ket{\delta a}
+ \bra{\delta a} \Sigma_{h} \ket{a}\,,
\end{align}
where $h = ({\rm N},{\rm O})$ and the
operators $\Sigma_{h} = \Sigma_{\rm N}$ and $\Sigma_{\rm O}$ are the
nested and overlapping irreducible operators, respectively, which appeared in
two-loop calculations for the Lamb shift. Furthermore,
$\ket{\delta a}$ is
the first-order perturbation of the reference-state wave function $\ket{a}$  by $V_g$,
\begin{align}\label{eq:pwf}
\ket{\delta a} = \sum_{n\ne a} \frac{\ket{n} \bra{n} V_g \ket{a}}{\vare_a-\vare_n}
\,.
\end{align}
The reducible contributions
will be referred to as the {\em derivative} (D) terms. They
have the following structure
\begin{align}
\label{eq:03}
E_{h{\rm D}} = \bra{a} V_g \ket{a}\, \bra{a} \frac{\partial \Sigma_h(\vare)}{\partial \vare} \ket{a}\Big|_{\vare = \vare_a}
\,.
\end{align}

Contributions of diagrams with the magnetic interaction attached
to the electron propagators are referred to as
the {\em vertex} (V) corrections. They are irreducible and have the
following structure
\begin{align}
\label{eq:04}
E_{h{\rm V}} = \bra{a} \Lambda_{h} \ket{a}\,,
\end{align}
where $h = ({\rm NL},{\rm NS},{\rm OL},{\rm OS})$.
$\Lambda_{\rm NL}$, $\Lambda_{\rm NS}$, $\Lambda_{\rm OL}$, and
$\Lambda_{\rm OS}$ are the nested (N) and overlapping (O) ladder (L) and side (S)
operators depicted in Fig.~\ref{fig:sese}(e), (f), (h), and (i), respectively.

Formal expressions for the SESE correction discussed so far suffer from
ultraviolet (UV) and infrared (IR) divergencies that need to be eliminated
before any actual numerical calculations can be performed. First we discuss
the renomalization and elimination of UV divergences.
Following the approach developed for the Lamb shift
\cite{mallampalli:98:pra,yerokhin:01:sese,yerokhin:03:prl,yerokhin:06:prl},
we base our renormalization procedure on separation of one or two first
terms of expansion of the Dirac-Coulomb propagator $G$
in the number of interactions
with the binding Coulomb field $V_C(r) = -\Za/r$. Specifically,
\begin{align}\label{eq:05}
G(\vare,\bfx_1,\bfx_2)  = &\ G^{(0)}(\vare,\bfx_1,\bfx_2)
 + G^{(1)}(\vare,\bfx_1,\bfx_2)
  \nonumber \\ &
 + G^{(2)}(\vare,\bfx_1,\bfx_2)  +
\ldots\,,
\end{align}
where the index $k$ in $G^{(k)}$ denotes the number of interactions with $V_C$.
The Dirac-Coulomb propagator is defined
as
\begin{equation}
G(\vare,\bfx_1,\bfx_2) = \sum_n \frac{\psi_n(\bfx_1)\,\psi_n^{\dag}(\bfx_2)}{\vare-\vare_n}\,,
\end{equation}
where $\psi_n(\bfx)$ are solutions of the Dirac equation with the Coulomb potential $V_C$,
$\vare_n$ are the corresponding energies, and
the sum over $n$ implies the summation over the full Dirac spectrum (i.e., summation over the principal quantum
number of the bound states, integration over the continuum states, summations over the angular-momentum quantum
number $\kappa_n$ and the angular momentum projections $\mu_n$).
Furthermore, $G^{(0)}$ is the free-electron propagator,
which is the limiting value of $G$ at $Z\to 0$. The free-electron propagator
is known both in coordinate as well as momentum space. In momentum representation,
it is just
\begin{align}\label{eq:06}
G(\vare,\bfp) = \frac{1}{\crossed{\rp}-m}\, \gamma^0 \, ,
\end{align}
where $\crossed{\rp} = \gamma_{\mu}p^{\mu}$ and $\rp = (\vare,\bfp)$.
The one-potential Green function
$G^{(1)}$ contains one interaction with the binding nuclear potential
and is given by
\begin{equation}
G^{(1)}(\vare,\bfx_1,\bfx_2) = \int d\bfx_3\,G^{(0)}(\vare,\bfx_1,\bfx_3)
 V_C(x_3) G^{(0)}(\vare,\bfx_3,\bfx_2)\,.
\end{equation}

The renormalization procedure at the two-loop level was developed for the Lamb shift in
Refs.~\cite{mallampalli:98:pra,yerokhin:01:sese,yerokhin:03:prl,yerokhin:06:prl}
and generalized for the $g$-factor in Refs.~\cite{sikora:18:phd,sikora:20}.
The resulting diagrams to be computed are divided into three classes:
(i) those calculated in coordinate space ($M$-term),
(ii) those calculated in the mixed momentum-coordinate representation ($P$-term) and
(iii) those computed in momentum space ($F$-term).
So, each of the nested and overlapping contributions is represented as a sum of the
$M$-, $P$-, and $F$-term parts,
\begin{align}\label{eq:08}
E_i = E_i^M + E_i^P + E_i^F\,.
\end{align}

In the present work, we will concentrate on the $P$-term contributions.
We define the $P$ term to be both UV and IR finite, so it can be computed separately from the remaining parts of the SESE
correction.

The main complication of the two-loop diagrams as compared to the one-loop case is
the presence of the overlapping UV divergences. E.g., the nested diagrams shown in
Fig.~\ref{fig:sese}({\em d})-({\em f}) can UV diverge due to the inner self-energy loop or due to the outer
self-energy loop, or both. This means that if we renormalize the inner loop, the overall
diagram will be still divergent due to the outer loop.
The distinct feature of the $P$-term contributions is that they contain an one-loop
UV-divergent subgraph (the free-electron self-energy or the free-electron
vertex), whereas the UV divergency of the overall diagram is cancelled by
suitably chosen subtractions.
The one-loop subgraph is renormalized and calculated in momentum space.
The outer diagram contains Dirac-Coulomb propagators which are known
in coordinate space.
As a compromise, we perform our computations in
the mixed momentum-coordinate representations, with the renomalized
one-loop subgraph computed in momentum space, whereas the outer diagram
is calculated in coordinate space. Each of the $P$-term contributions
contains one unbounded partial-wave expansion, which originates from Dirac-Coulomb propagators.
\\

\section{Perturbed-orbital contributions}

\begin{figure*}
\centerline{
\resizebox{0.75\textwidth}{!}{%
  \includegraphics{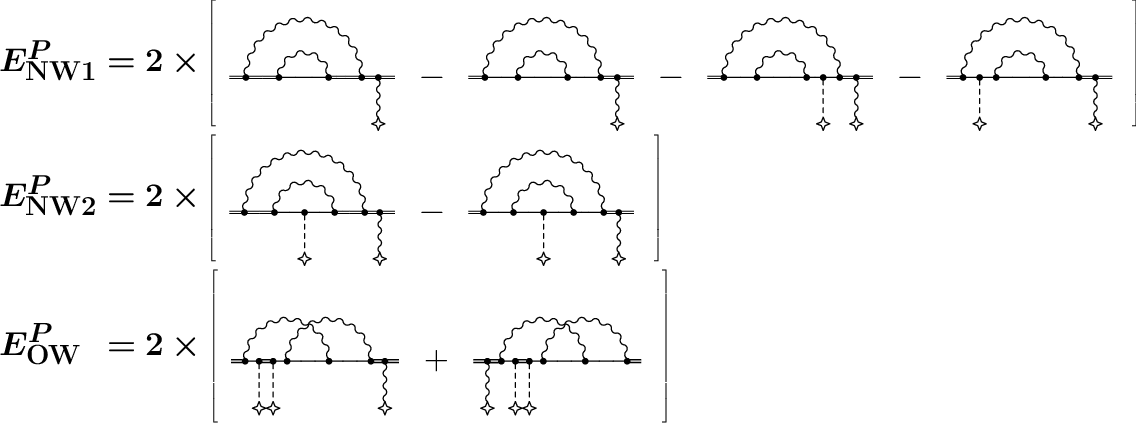}
}
}
\caption{
Diagrammatic representation of the perturbed-orbital $P$-term contributions.
A double line denotes the bound electron propagator, the single line denotes the
free electron propagator, the wave line denotes the photon propagator,
the dashed line terminated by a stylized cross denotes the
interaction with the Coulomb nuclear field, the wave line terminated by
the stylized cross denotes the interaction with the external magnetic field.
\label{fig:Ppo}
}
\end{figure*}

The perturbed-orbital $P$-term contributions are diagrammatically represented in Fig.~\ref{fig:Ppo}.
They are related to the corresponding contributions
to the Lamb shift that appeared in our previous investigations
\cite{yerokhin:10:sese,yerokhin:18:sese}. Their expressions can be obtained by perturbing
the external reference-state wave functions in the Lamb-shift expressions
by the magnetic potential,
$\psi_{a} \to \psi_{\delta a}$,
where $\psi_{\delta a}$ is given by Eq.~(\ref{eq:pwf}).

\begin{widetext}

\subsection{$\bm{E^{P}_{\rm NW1}}$}
This correction can be obtained from the $N1$ $P$-term contribution to the Lamb shift
[see Eq.~(113) of Ref.~\cite{yerokhin:03:epjd}] by perturbing the external wave functions,
$\psi_a \to \psi_{\delta a}$
(the first line of Fig.~\ref{fig:Ppo}). The result is 
\begin{align}\label{eq:NW1}
E^P_{\mathrm{NW1}} = &\,
    2i\alpha \int_{C_F} d\omega
    \int \frac{d\bfp}{(2\pi)^3} \int d\bfx_1d\bfx_2\,
    D(\omega,x_{12})\,\psi_a^{\dag}(\bfx_1)\, \alpha_{\mu}
    \bigg\{
        G(E,\bfx_1,\bfp)\, \gamma^0 \Sigma^{(0)}_R(E,\bfp)\, G(E,\bfp,\bfx_2)
\nonumber \\ &
     -
        G^{(0)}(E,\bfx_1,\bfp)\, \gamma^0 \Sigma^{(0)}_R(E,\bfp)\, G^{(0)}(E,\bfp,\bfx_2)
     -
        G^{(1)}(E,\bfx_1,\bfp)\, \gamma^0 \Sigma^{(0)}_R(E,\bfp)\, G^{(0)}(E,\bfp,\bfx_2)
\nonumber \\ &
     -
        G^{(0)}(E,\bfx_1,\bfp)\, \gamma^0 \Sigma^{(0)}_R(E,\bfp)\, G^{(1)}(E,\bfp,\bfx_2)
     \bigg\}
       \alpha^{\mu}  \psi_{\delta a}(\bfx_2) + (\psi_a \leftrightarrow \psi_{\delta a})\,,
\end{align}
where $E \equiv \vare_a - \omega$, $D(\omega,x_{12})$ is the scalar part of the photon propagator in the
Feynman gauge,
\begin{equation}
D(\omega,x_{12}) = \frac{\exp(i\, \sqrt{\omega^2+i0}\,x_{12})}{4\pi x_{12}}
\,,
\end{equation}
where the branch of the square root is fixed by the condition ${\rm
Im}(\sqrt{\omega^2+i0})>0$, and $x_{12} = |\bfx_1-\bfx_2|$.
Furthermore, $\Sigma^{(0)}_R$ is the renormalized free-electron self-energy
operator in $D = 4$ dimensions, see Eq.~(241) of Ref.~\cite{yerokhin:03:epjd},
$C_F$ is the standard Feynman integration contour,
and $\psi_{\delta a}$ is the perturbed reference-state wave function (\ref{eq:pwf}).
The Dirac-Coulomb Green function in the mixed coordinate-momentum representation is
defined as a Fourier transform of the coordinate-space Dirac-Coulomb Green function
over one of the radial variables \cite{yerokhin:03:epjd},
\begin{equation} \label{eq3}
G(\vare,\bfx_1,\bfp) =
        \int d\bfx_2 \, e^{i\bfp\cdot \bfx_2} \,
       G(\vare,\bfx_1,\bfx_2)          \,, \ \
G(\vare,\bfp,\bfx_2) =
        \int d\bfx_1 \, e^{-i\bfp\cdot \bfx_1}\,
          G(\vare,\bfx_1,\bfx_2) \, .
\end{equation}

For convenience of the numerical evaluation, the integrand in curly brackets in Eq.~(\ref{eq:NW1})
was transformed as \cite{yerokhin:10:sese}
\begin{align}\label{eq:NW1:2}
\Big\{ \ldots \Big\}
   = &\
     G_V(E,\bfx_1,\bfp)\, \frac1{\slashed{p}-m}\, \Sigma^{(0)}_R(E,\bfp)\, \frac1{\slashed{p}-m}\,
       G_V(E,\bfp,\bfx_2)
  \nonumber \\
 &    +2\,
      \Big[  G_V(E,\bfx_1,\bfp) - G^{(0)}_V(E,\bfx_1,\bfp)\Big]
         \frac1{\slashed{p}-m}\, \Sigma^{(0)}_R(E,\bfp)\, G^{(0)}(E,\bfp,\bfx_2)
\,,
\end{align}
where $\slashed{p} = \gamma^0E - \bgamma\cdot\bfp$ and $G_V$ denotes
the Fourier transform of
the product of the Green function and the Coulomb potential
\cite{yerokhin:10:sese},
\begin{equation}
G_V(\vare,\bfx_1,\bfp) =
        \int d\bfx_2 \, e^{i\bfp\cdot \bfx_2} \,
       G(\vare,\bfx_1,\bfx_2) \, V_C(x_2)
         \,, \ \
G_V(\vare,\bfp,\bfx_2) =
        \int d\bfx_1 \, e^{-i\bfp\cdot \bfx_1}\,
    V_C(x_1)\,
          G(\vare,\bfx_1,\bfx_2) \, .
\end{equation}
The last term in Eq.~(\ref{eq:NW1:2}) contains a combinatorial factor of 2 that accounts for
the contributions of two equivalent diagrams. Note that these two diagrams become equivalent
only after the symmetrization $\psi_a \leftrightarrow \psi_{\delta a}$.
The idea behind the transformation (\ref{eq:NW1:2}) is that it separates out the additional
free-electron propagator(s), which decrease  as $\sim 1/p$ for large momenta.
This separation simplifies the numerical integration over $p$ since it makes the
integrand to behave better for large $p$.
It should be noted that, contrary to the Lamb-shift case, $E^P_{\mathrm{NW1}}$ does not contain
IR divergences.
The reason is that the reference-state contribution in the IR
(i.e., small-$\omega$) region is regularized by the
orthogonality of $\psi_{\delta a}$ and $\psi_{a}$, $\lbr \delta a | a\rbr = 0$.
For this reason, there are no infrared subtractions in $E^P_{\mathrm{NW1}}$.
We compute $E^P_{\mathrm{NW1}}$ by generalizing the code developed for the Lamb shift and
described in our previous publications \cite{yerokhin:10:sese,yerokhin:18:sese}.

\subsection{$\bm{E^{P}_{\rm NW2}}$}
This correction can be obtained from the $N2$ $P$-term contribution to the Lamb shift
[see Eq.~(117) of Ref.~\cite{yerokhin:03:epjd}] by perturbing the external wave functions,
$\psi_a \to \psi_{\delta a}$ (the second line of Fig.~\ref{fig:Ppo}). The result is 
\begin{align}\label{eq:NW2}
E^P_{\mathrm{NW2}} = &\,
    2i\alpha \int_{C_F} d\omega
    \int \frac{d\bfp_1}{(2\pi)^3} \frac{d\bfp_2}{(2\pi)^3} \int d\bfx_1d\bfx_2\,
    D(\omega,x_{12})\,V_C(\bfq)\, \psi_a^{\dag}(\bfx_1)\, \alpha_{\mu}
    \bigg\{
        G(E,\bfx_1,\bfp_1)\, \gamma^0 \Gamma(E,\bfp_1;E,\bfp_2)\, G(E,\bfp_2,\bfx_2)
\nonumber \\ &
     \ \ \ \ \ -
        G^{(0)}(E,\bfx_1,\bfp_1)\, \gamma^0 \Gamma(E,\bfp_1;E,\bfp_2)\, G^{(0)}(E,\bfp_2,\bfx_2)
     \bigg\}
       \alpha^{\mu}  \psi_{\delta a}(\bfx_2) + (\psi_a \leftrightarrow \psi_{\delta a})\,,
\end{align}
where $\Gamma(E,\bfp_1;E,\bfp_2)$ is the time component
of the renormalized one-loop vertex operator in $D = 4$ dimensions, $\Gamma \equiv \Gamma^0_R$ (for brevity we will suppress
the indices here), see Eq.~(257) of Ref.~\cite{yerokhin:03:epjd} and Appendix~B of Ref.~\cite{yerokhin:99:pra}.
For convenience of
the numerical evaluation, the integrand in curly brackets in Eq.~(\ref{eq:NW2}) is transformed as \cite{yerokhin:10:sese}
\begin{align}\label{eq:NW2:2}
\Big\{ \ldots \Big\}
   = &\
     G_V(E,\bfx_1,\bfp_1)\, \frac1{\slashed{p}_1-m}\, \Gamma(E,\bfp_1;E,\bfp_2)\, \frac1{\slashed{p}_2-m}\,
       G_V(E,\bfp_2,\bfx_2)
  \nonumber \\
 &    +2\,
      G_V(E,\bfx_1,\bfp_1)
         \frac1{\slashed{p}_1-m}\, \Gamma(E,\bfp_1;E,\bfp_2)\, G^{(0)}(E,\bfp_2,\bfx_2)
\,,
\end{align}
where $\slashed{p}_{1,2} = \gamma^0E - \bgamma\cdot\bfp_{1,2}$ and the combinatorial factor of 2 accounts for
two equivalent diagrams.

Equation (\ref{eq:NW2}) contains two integrations over momenta, $\bfp_1$ and $\bfp_2$.
Its numerical evaluation is complicated by the presence of the (integrable) Coulomb singularity
at $\bfp_1 = \bfp_2$ coming through the Coulomb potential $V_C(\bfq)$.
It is advantageous to
separate out the diagonal in momentum contribution of the vertex operator and evaluate it separately.
Therefore, we split the vertex operator into the diagonal and the nondiagonal (in momentum) parts,
\begin{align} \label{eq:vertex_separation}
\Gamma(E,\bfp_1;E,\bfp_2) = &\,
\Gamma_{\rm dia}(E,\bfp_1) + \Gamma_{\rm ndia}(E,\bfp_1;E,\bfp_2)
 \equiv
\Gamma(E,\bfp_1;E,\bfp_1) + \Big[ \Gamma(E,\bfp_1;E,\bfp_2)
 - \Gamma(E,\bfp_1;E,\bfp_1)\Big]\,.
\end{align}
In the diagonal part, the integration over one momentum (and the Coulomb singularity) is carried
out analytically by using the
identity
\begin{align}
\int \frac{d\bfp_2}{(2\pi)^3}\, V_C(\bfp_1-\bfp_2)\, G(\bfp_2,\bfx_2) = G_V(\bfp_1,\bfx_2)\,.
\end{align}
Eq.~(\ref{eq:NW2}) is very similar to the corresponding part of the $P$-term for the Lamb shift.
An important difference is that $E^P_{\mathrm{NW2}}$, similarly to $E^P_{\mathrm{NW1}}$ and contrary
to the Lamb-shift case,
does not contain IR divergences,
because of the orthogonality of $\psi_a $ and $\psi_{\delta a}$.
We compute this contribution by generalizing the code developed for the Lamb shift and
described in our previous studies \cite{yerokhin:10:sese,yerokhin:18:sese}.

\subsection{$\bm{E^{P}_{\rm OW}}$}
This correction can be obtained from the overlapping $P$-term contribution to the Lamb shift
[see Eq.~(120) of Ref.~\cite{yerokhin:03:epjd}] by perturbing the external wave functions,
$\psi_a \to \psi_{\delta a}$
(see the third line of Fig.~\ref{fig:Ppo}). The result is
\begin{align}
E^P_{\mathrm{OW}} =
    -4i\alpha \int_{C_F} d\omega
    &\,
    \int \frac{d\bfp_1}{(2\pi)^3} \frac{d\bfp_2}{(2\pi)^3} \int d\bfz\,
    \frac{e^{-i\bfq\cdot\bfz}}{\omega^2-\bfq\,^2+i0}\, \psi_a^{\dag}(\bfx_1)\, \alpha_{\mu}
\nonumber \\ & \times
        G^{(2+)}(E,\bfx_1,\bfp_1)\, \gamma^0 \Gamma^{\mu}_R(E,\bfp_1;\vare_a,\bfp_2)\,
        \psi_{\delta a}(\bfp_2) + (\psi_a \leftrightarrow \psi_{\delta a})\,,
\end{align}
where $G^{(2+)} = G - G^{(0)}- G^{(1)}$ is the electron propagator with two or more
interactions with the binding potential, and
$\Gamma^{\mu}_R$ is the renormalized free-electron vertex operator in $D = 4$ dimensions,
see Eq.~(257) of Ref.~\cite{yerokhin:03:epjd} and Appendix~B of Ref.~\cite{yerokhin:99:pra}.
The computation of this contribution was performed as described in Ref.~\cite{yerokhin:18:sese},
after transforming the integrand as follows
\begin{align}
        G^{(2+)}(E,\bfx_1,\bfp_1)\, \gamma^0 \Gamma^{\mu}_R(E,\bfp_1;\vare_a,\bfp_2)
        =
        \Big[
        G_V(E,\bfx_1,\bfp_1)
        -
        G^{(0)}_V(E,\bfx_1,\bfp_1)
        \Big] \,
           \frac1{\slashed{p}_1-m}\,
                \Gamma^{\mu}_R(E,\bfp_1;\vare_a,\bfp_2)\,.
\end{align}

\section{Derivative contributions}

\begin{figure*}
\centerline{
\resizebox{0.6\textwidth}{!}{%
  \includegraphics{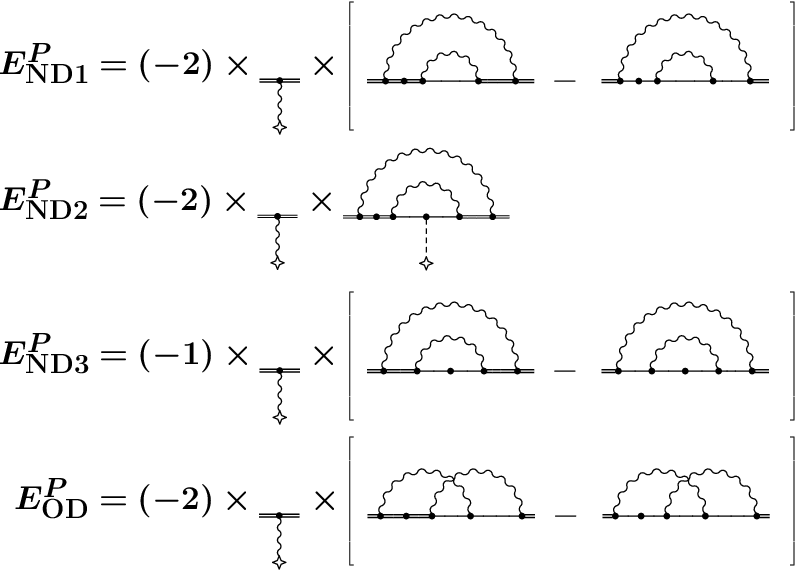}
}
}
\caption{
Diagrammatic representation of the derivative $P$-term contributions.
\label{fig:Pderiv}
}
\end{figure*}

The derivative $P$-term contributions are shown in Fig.~\ref{fig:Pderiv}.
They are related
to the $P$-term contributions for
the Lamb shift. As compared to their Lamb-shift counterparts, they contain a derivative of one of the
electron propagators over the energy argument. The derivative is equivalent to the insertion of
a scalar vertex into the corresponding Feynman diagram,
\begin{align}
 \frac{\partial}{\partial \vare}\, G(\vare,\bfx_1,\bfx_2) =
  - \int d\bfx_3\, G(\vare,\bfx_1,\bfx_3)\, G(\vare,\bfx_3,\bfx_2)\,.
\end{align}
The additional vertex in the diagrams not only makes the numerical evaluation more difficult than in the Lamb-shift case,
but also enhances the degree of IR divergences.

\subsection{$\bm{E^{P}_{\rm ND1}}$}

The first nested derivative contribution is induced by the first line of Fig.~\ref{fig:Pderiv}.
Its expression is given by
\begin{align}\label{eq:nd1}
E^P_{\mathrm{ND1}} = & (-2)\, \lbr V_g \rbr\,
    2i\alpha \int_{C_F} d\omega
    \int \frac{d\bfp}{(2\pi)^3} \int d\bfx_1d\bfx_2d\bfx_3\,
    D(\omega,x_{12})\,\psi_a^{\dag}(\bfx_1)\, \alpha_{\mu}
\nonumber \\ & \times
    \Big[
        G(E,\bfx_1,\bfx_3)\, G(E,\bfx_3,\bfp)\, \gamma^0 \Sigma^{(0)}_R(E,\bfp)\, G(E,\bfp,\bfx_2)
\nonumber \\ &
 \ \ \
     -
        G^{(0)}(E,\bfx_1,\bfx_3)\, G^{(0)}(E,\bfx_3,\bfp)\, \gamma^0 \Sigma^{(0)}_R(E,\bfp)\, G^{(0)}(E,\bfp,\bfx_2)
     \Big]
       \alpha^{\mu} \psi_a(\bfx_2)\,,
\end{align}
where $\lbr V_g \rbr$ is the expectation value of the effective $g$-factor operator given by
Eq.~(\ref{eq:01}) and
$\Sigma^{(0)}_R(E,\bfp)$ is the renormalized free-electron self-energy operator.
In order to separate the infrared divergences present in the above expression, we subtract and re-add the
following terms in the expression in the brackets,
\begin{align}\label{eq:nd1s}
\Big[ \ldots \Big] = &\, \Big[ \ldots
     \mp
        G^{(a)}(E,\bfx_1,\bfx_3)\, G(E,\bfx_3,\bfp)\, \gamma^0 \Big(
                    \Sigma^{(0)}_R(\vare_a,\bfp)-\omega\, \Sigma^{(0)^{\prime}}_R(\vare_a,\bfp)\Big)
                                                    \, G^{(a)}(E,\bfp,\bfx_2) \Big]\,,
\end{align}
where $ \Sigma^{(0)^{\prime}}_R(\vare_a,\bfp) = d/(d\vare)  \Sigma^{(0)}_R(\vare,\bfp)|_{\vare = \vare_a}$
and $G^{(a)}$ is the reference-state part of the Dirac-Coulomb propagator defined as
\begin{equation}
G^{(a)}(\vare,\bfx_1,\bfx_2) = \sum_{\mu_{a'}} \frac{\psi_{a'}(\bfx_1)\,\psi_{a'}^{\dag}(\bfx_2)}{\vare-\vare_a}\,,
\end{equation}
where $a'$ denotes the electron state that differs from the reference-state $a$ only by the angular-momentum projection $\mu_{a'}$.
The sign $\mp$ in Eq.~(\ref{eq:nd1s})
means that we subtract the infrared-divergent part and then re-add it and evaluate separately. This leads
to a separation of $E^P_{\mathrm{ND1}}$ into the regular (R) and two infrared-divergent
($\mathrm{IR'}$, IR) contributions,
\begin{align}
   E^P_{\mathrm{ND1}} = E^P_{\mathrm{ND1}}(\mathrm{R}) + E^P_{\mathrm{ND1}}(\mathrm{IR'}) + E^P_{\mathrm{ND1}}(\mathrm{IR})\,.
\end{align}
The infrared IR and IR$'$ contributions are regularized by introducing a fictitious finite photon mass
$\mu$; the $\omega$ integration in them is evaluated analytically.
After IR-divergent contributions are added together, the $\mu$ divergence disappears and
the limit $\mu\to 0$ can be taken.
Here and in what follows we will denote the linearly-divergent ($\sim 1/\mu$) infrared contribution as
$\mathrm{IR'}$, and the logarithmically-divergent ($\sim \ln\mu$) infrared contribution as IR.
Noting that in Eq.~(\ref{eq:nd1s}) the free-electron operators do not
depend on $\omega$, we factorize out the infrared terms and obtain
\begin{align}
E^P_{\mathrm{ND1}}(\mathrm{IR'}) = (-2)\,\lbr V_g \rbr\,\big<  \gamma^0\Sigma^{(0)}_R(\vare_a)\big>\, J_3\,,
\ \ \
E^P_{\mathrm{ND1}}(\mathrm{IR}) = (-2)\,\lbr V_g \rbr\,\big<  \gamma^0\Sigma^{(0)^{\prime}}_R(\vare_a)\big>\, J_2\,.
\end{align}
Here, $J_2$ and $J_3$ are
basic IR-divergent integrals defined as (see Appendix~B of Ref.~\cite{yerokhin:20:green})
\begin{align}\label{eq:Ja}
J_{\alpha} = \sum_{\mu_{a'}}\frac{i}{2\pi} \int_{C_F} d\omega\, \frac{\lbr aa'|I_{\mu}(\omega)|a'a\rbr}
  {(-\omega+i0)^{\alpha}}\,,
\end{align}
where
$I_{\mu}(\omega)$ is the electron-electron interaction operator with a finite photon mass $\mu$,
\begin{align}
I_{\mu}(\omega,x_{12}) = \alpha\,\alpha_{\sigma}\alpha^{\sigma}\,
\frac{e^{i\sqrt{\omega^2-\mu^2+i0}\,x_{12}}}{x_{12}}\,.
\end{align}
The IR divergent integrals are evaluated as \cite{yerokhin:20:green}
\begin{align}\label{eq:Jb}
J_{2} = &\
  \frac{\alpha}{\pi} \bigg[
  \ln \frac{\mu}{2}+\gamma
  + \sum_{\mu_{a'}} \bra{aa'}\alpha_{\sigma}\alpha^{\sigma}\,\ln x_{12}\ket{a'a}
  \bigg]\,,
  \ \ \
J_{3} =
  \frac{\alpha}{4} \bigg[
  \frac1{\mu}
  - \sum_{\mu_{a'}} \bra{aa'}\alpha_{\sigma}\alpha^{\sigma}\,x_{12}\ket{a'a}
  \bigg]\,.
\end{align}

For the numerical evaluation it is convenient to transform the integrand of Eq.~(\ref{eq:nd1})
to a form that does not involve subtraction of the free-electron contribution. We thus write the
integrand as
\begin{align}\label{eq:nd1:4}
\int d\bfx_3\Big[
        G(E,\bfx_1,\bfx_3)\, G(E,\bfx_3,\bfp)\, \gamma^0 & \, \Sigma^{(0)}_R(E,\bfp)\, G(E,\bfp,\bfx_2)
     -
        G^{(0)}(E,\bfx_1,\bfx_3)\, G^{(0)}(E,\bfx_3,\bfp)\, \gamma^0 \Sigma^{(0)}_R(E,\bfp)\, G^{(0)}(E,\bfp,\bfx_2) \Big]
 =
  \nonumber \\
 &   =
     G_V(E,\bfx_1,\bfp)\, \frac1{\slashed{p}-m}\, \gamma^0 \frac1{\slashed{p}-m} \,\Sigma^{(0)}_R(E,\bfp)\, G^{(0)}(E,\bfp,\bfx_2)\,
  \nonumber \\
  &
  +
        G(E,\bfx_1,\bfp)\, \frac1{\slashed{p}-m}\, \Sigma^{(0)}_R(E,\bfp)\,\frac1{\slashed{p}-m}\, \gamma^0\, G_V(E,\bfp,\bfx_2)\,
  \nonumber \\
  &
  +
\int d\bfx_3\,
        G(E,\bfx_1,\bfx_3)\, G_V(E,\bfx_3,\bfp)\, \frac1{\slashed{p}-m} \,\Sigma^{(0)}_R(E,\bfp)\, G(E,\bfp,\bfx_2)
\,,
\end{align}
with $\slashed{p} = \gamma^0E - \vec{\gamma}\cdot\bfp$. An advantage of the representation (\ref{eq:nd1:4})
is that each of the three terms in the right-hand-side contains at least one Coulomb interaction explicitly.
A useful check for the numerical procedure
for $E_{\mathrm{ND1}}^{{P}}$ can be obtained by making the replacement
$\gamma^0 \Sigma^{(0)}_R(\vare,\bfp) \to 1$, after which the contribution can be computed entirely
in coordinate
representation. Specifically, after this replacement
the regular part of $E_{\mathrm{ND1}}^{\mathrm{P}}$ becomes
\begin{align}
E_{\mathrm{ND1}}^{{P}}(\mathrm{R})  \to &\, (-2) \, \lbr V_g \rbr\,
    2i\alpha \int_{C_F} d\omega
    \int d\bfx_1d\bfx_2\,
    D(\omega,x_{12})\,\psi_a^{\dag}(\bfx_1)\, \alpha_{\mu}
   \nonumber \\ & \times
    \, \frac12 \frac{\partial^2}{(\partial E)^2}
      \Big[G(E,\bfx_1,\bfx_2)- G^{(0)}(E,\bfx_1,\bfx_2) - G^{(a)}(E,\bfx_1,\bfx_2)\Big]
             \alpha^{\mu} \psi_a(\bfx_2)\,,
\end{align}
which can be computed in a straightforward manner in coordinate space.

\subsection{$\bm{E^{P}_{\rm ND2}}$}
The second nested derivative contribution is induced by the second line of Fig.~\ref{fig:Pderiv}
and is expressed as
\begin{align}\label{eq:nd2}
E^P_{\mathrm{ND2}} = & (-2)\, \lbr V_g \rbr\,
    2i\alpha \int_{C_F} d\omega
    \int \frac{d\bfp_1}{(2\pi)^3} \frac{d\bfp_2}{(2\pi)^3} \int d\bfx_1d\bfx_2d\bfx_3\,
    D(\omega,x_{12})\,V_C(q)\, \psi_a^{\dag}(\bfx_1)\, \alpha_{\mu}
\nonumber \\ & \times
    \Big[
        G(E,\bfx_1,\bfx_3)\, G(E,\bfx_3,\bfp_1) \,\gamma^0\, \Gamma(E,\bfp_1;E,\bfp_2)\, G(E,\bfp_2,\bfx_2)
     \Big]
       \alpha^{\mu} \psi_a(\bfx_2)\,,
\end{align}
where 
$\Gamma(\vare_1,\bfp_1;\vare_2,\bfp_2)$ is the time component of the renormalized
free-electron vertex operator.
In order to separate out the IR divergences present in the above expression, we subtract and then re-add the following terms
in the brackets above
\begin{align}\label{eq:nd2:2}
\Big[ \ldots \Big] = &\, \Big[ \ldots
     \mp
        G^{(a)}(E,\bfx_1,\bfx_3)\, G(E,\bfx_3,\bfp_1)\, \gamma^0  \Big(
                    \Gamma(\vare_a,\bfp_1;\vare_a,\bfp_2)
                    -\omega\, \Gamma^{\prime}(\vare_a,\bfp_1;\vare_a,\bfp_2)
                                    \Big)
                                                    \, G^{(a)}(E,\bfp_2,\bfx_2) \Big]\,,
\end{align}
where $\Gamma^{\prime}(\vare_a,\bfp_1;\vare_a,\bfp_2) = d/(d\vare) \Gamma(\vare,\bfp_1;\vare,\bfp_2)|_{\vare = \vare_a}$.
This separates $E^P_{\mathrm{ND2}}$ into a regular and two IR-divergent terms,
\begin{align}
E^P_{\mathrm{ND2}} = E^P_{\mathrm{ND2}}(\mathrm{R}) + E^P_{\mathrm{ND2}}(\mathrm{IR'}) + E^P_{\mathrm{ND2}}(\mathrm{IR})\,.
\end{align}
The infrared parts can be factorized and expressed as
\begin{align}
E^P_{\mathrm{ND2}}(\mathrm{IR'}) = (-2)\,\lbr V_g \rbr\,\big< V_C\, \gamma^0 \,\Gamma(\vare_a;\vare_a) \big>\, J_3\,,
\\
E^P_{\mathrm{ND2}}(\mathrm{IR}) = (-2)\,\lbr V_g \rbr\,\big< V_C\, \gamma^0 \,\Gamma^{\prime}(\vare_a;\vare_a)\big>\, J_2\,,
\end{align}
with integrals $J_2$ and $J_3$ defined by Eq.~(\ref{eq:Ja}) and evaluated in Eqs.~(\ref{eq:Jb}).
The numerical evaluation of the regular (R) part is complicated by the presence of
the Coulomb singularity at $q = 0$. This problem is handled by the splitting the
vertex operator into the diagonal and the non-diagonal (in momentum) parts, according to
Eq.~(\ref{eq:vertex_separation}).
The diagonal part is re-written as
\begin{align}\label{eq:2b}
E^P_{\mathrm{ND2}}(\mathrm{R,diag}) = 2\, \lbr V_g \rbr\,
    2i\alpha \int_{C_F} d\omega
     & \,
    \int \frac{d\bfp}{(2\pi)^3} \int d\bfx_1d\bfx_2d\bfx_3\,
    D(\omega,x_{12})\,\psi_a^{\dag}(\bfx_1)\, \alpha_{\mu}
\nonumber \\ & \times
        G(E,\bfx_1,\bfx_3)\, G(E,\bfx_3,\bfp)\, \gamma^0\, \Sigma^{(0)^{\prime}}_R(E,\bfp)\, G_V(E,\bfp,\bfx_2)
       \alpha^{\mu} \psi_a(\bfx_2)\,.
\end{align}
This expression contains only one integration over the momentum and its evaluation is very similar to that
for the $E^P_{\mathrm{ND1}}$ term.
In the nondiagonal part, the integrand is a smooth function of $\bfp_1$ and
$\bfp_2$ and the momentum integrations are much easier to calculate numerically.
Another important advantage is that the partial-wave expansion of the
nondiagonal part converges much faster than that
of the diagonal term, so we need to compute less (typically, about 10)
partial waves, whereas for the diagonal term we extended the partial-wave expansion
up to $|\kappa_{\rm max}| = 25$-30.

\subsection{$\bm{E^{P}_{\rm ND3}}$}
The third nested derivative contribution is induced by the third line of Fig.~\ref{fig:Pderiv}.
Its expression is given by
\begin{align}\label{eq:nd3}
E^P_{\mathrm{ND3}} = &\, \lbr V_g \rbr\,
    2i\alpha \int_{C_F} d\omega
    \int \frac{d\bfp}{(2\pi)^3} \int d\bfx_1d\bfx_2\,
    D(\omega,x_{12})\,\psi_a^{\dag}(\bfx_1)\, \alpha_{\mu}
\nonumber \\ & \times
    \Big[
        G(E,\bfx_1,\bfp)\, \gamma^0 \Sigma^{(0)^{\prime}}_R(E,\bfp)\, G(E,\bfp,\bfx_2)
     -
        G^{(0)}(E,\bfx_1,\bfp)\, \gamma^0 \Sigma^{(0)^{\prime}}_R(E,\bfp)\, G^{(0)}(E,\bfp,\bfx_2)
     \Big]
       \alpha^{\mu} \psi_a(\bfx_2)\,.
\end{align}
In order to separate out the IR divergence, we subtract and then re-add the following
contribution in the brackets,
\begin{align}
\Big[ \ldots \Big] = &\, \Big[ \ldots
     \mp
        G^{(a)}(E,\bfx_1,\bfp)\, \gamma^0 \Sigma^{(0)^{\prime}}_R(\vare_a,\bfp)\, G^{(a)}(E,\bfp,\bfx_2) \Big]\,.
\end{align}
This transformation separates $E^P_{\mathrm{ND3}}$ into the regular and the infrared parts,
\begin{align}
E^P_{\mathrm{ND3}}
      =  E^P_{\mathrm{ND3}}(\mathrm{R}) + E^P_{\mathrm{ND3}}(\mathrm{IR})\,.
\end{align}
The infrared part is expressed as
\begin{align}
E^P_{\mathrm{ND3}}(\mathrm{IR}) = \lbr V_g \rbr\,\big<  \gamma^0\Sigma^{(0)^{\prime}}_R(\vare_a)\big>\, J_2\,.
\end{align}
For the numerical evaluation of the regular part, we transform the integrand as
\begin{align}
        G(E,\bfx_1,\bfp)\, \gamma^0 & \, \Sigma^{(0)^{\prime}}_R(E,\bfp)\, G(E,\bfp,\bfx_2)
     -
        G^{(0)}(E,\bfx_1,\bfp)\, \gamma^0 \Sigma^{(0)^{\prime}}_R(E,\bfp)\, G^{(0)}(E,\bfp,\bfx_2)
        + (\bfx_1 \leftrightarrow \bfx_2)
 =
  \nonumber \\
 &   =
     G_V(E,\bfx_1,\bfp)\, \frac1{\slashed{p}-m}\, \Sigma^{(0)^{\prime}}_R(E,\bfp)\,
      \Big[ G^{(0)}(E,\bfp,\bfx_2) + G(E,\bfp,\bfx_2)\Big] + (\bfx_1 \leftrightarrow \bfx_2)
\,.
\end{align}

\subsection{$\bm{E^{P}_{\rm OD}}$}

The overlapping derivative $P$-term contribution is represented by the bottom line of Fig.~\ref{fig:Pderiv}.
It is given by
\begin{align}
E^P_{\mathrm{OD}} = & \
    4i\alpha \,\lbr V_g\rbr\, \int_{C_F} d\omega
    \int \frac{d\bfp_1}{(2\pi)^3} \frac{d\bfp_2}{(2\pi)^3} \int d\bfz\, d\bfy\,
    \frac{e^{-i\bfq\cdot\bfz}}{\omega^2-\bfq\,^2 + i0}\,
    \psi_a^{\dag}(\bfz)\, \alpha_{\mu}
\nonumber \\ & \times
    \Big[
        G(E,\bfz,\bfy)\, G(E,\bfy,\bfp_1)
    - G^{(0)}(E,\bfz,\bfy)\, G^{(0)}(E,\bfy,\bfp_1)
    \Big]
         \gamma^0 \Gamma^{\mu}_R(E,\bfp_1;\vare_a,\bfp_2)\, \psi_a(\bfp_2)
         \,.
\end{align}
In order to remove the IR divergency, we subtract and re-add the following contribution in the
brackets:
\begin{align}
\Big[ \ldots \Big] = \Big[ \ldots  \mp
   G^{(a)}(E,\bfz,\bfy)\, G^{(a)}(E,\bfy,\bfp_1)\Big]\,.
\end{align}
This
separates $E^P_{\mathrm{OD}}$ into the regular (R) and infrared (IR) parts,
\begin{align}
E^P_{\mathrm{OD}} = E^P_{\mathrm{OD}}(\mathrm{R}) + E^P_{\mathrm{OD}}(\mathrm{IR})\,.
\end{align}
In the regular contribution, we perform the Wick rotation of the $\omega$ integration contour
as follows
\begin{align}
i \sum_{n \neq a}\int_{C_F} d\omega\, \frac{f_{n}(\omega)}{(\vare_a-\omega-u\vare_{n})^2}
 = -2\,  {\rm \Re}\, \sum_{n} \int_0^{\infty}d\omega\,
 \frac{f_{n}(i\omega)}{(\Delta_{an}-i\omega)^2}
 \,,
\end{align}
where $\Delta_{an} = \vare_a - \vare_n$.
The infrared contribution is given by
\begin{align}
E^P_{\mathrm{OD}}({\mathrm{IR}}) = & \
    4i\alpha\, \lbr V_g\rbr\,
    \int_{C_F} d\omega
    \frac{1}{(-\omega + i0)^2}
    \int \frac{d\bfp_1}{(2\pi)^3} \frac{d\bfp_2}{(2\pi)^3}
\nonumber \\ & \times
    \bigg(
     \int d\bfz
    \frac{e^{-i\bfq\cdot\bfz}}{\omega^2-\bfq\,^2+i0}\,
    \psi_a^{\dag}(\bfz)\, \alpha_{\mu} \,\psi_{a}(\bfz)\,
     \bigg)\,
        \psi_{a}^{\dag}(\bfp_1)\,
         \gamma^0 \Gamma^{\mu}_R(E,\bfp_1;\vare_a,\bfp_2)\, \psi_a(\bfp_2)
         \,.
\end{align}
We observe that the $\omega$ integration in the above expression is logarithmically divergent
at $\omega \to 0$. This divergence cancels out when $E^P_{\mathrm{OD}}({\mathrm{IR}})$ is considered
together with the corresponding contribution from the overlapping vertex
part, so we postpone its further evaluation until the vertex term is analysed.

\section{Vertex contributions}

The vertex $P$-term contributions are the most complicated ones since they have an additional interaction
inserted in the self-energy loop, as compared to the Lamb-shift case. The diagrammatic representation
of the vertex $P$-term contribution is presented in Fig.~\ref{fig:Pvertex}.

\begin{figure*}
\centerline{
\resizebox{0.6\textwidth}{!}{%
  \includegraphics{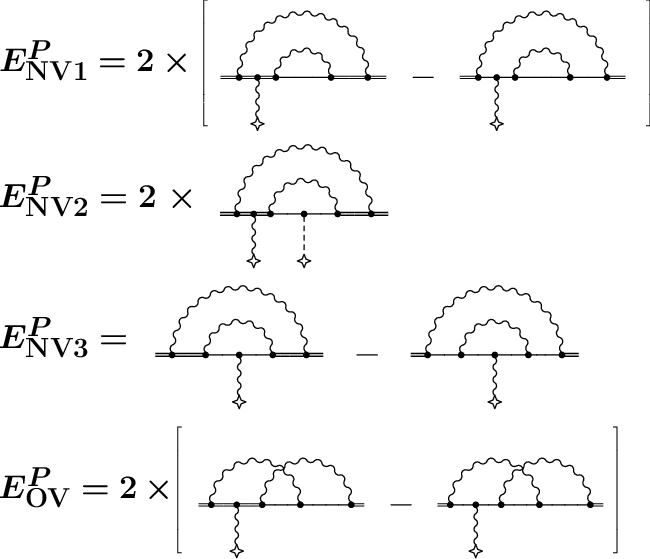}
}
}
\caption{
Diagrammatic representation of the vertex $P$-term contributions.
\label{fig:Pvertex}
}
\end{figure*}

\subsection{$\bm{E^{P}_{\rm NV1}}$}

The first nested vertex $P$-term contribution is diagrammatically represented
by the uppermost line of Fig.~\ref{fig:Pvertex}.
It is given by the expression
\begin{align}
E^P_{\mathrm{NV1}} = & \
    4i\alpha \int_{C_F} d\omega
    \int \frac{d\bfp}{(2\pi)^3} \int d\bfx_1d\bfx_2d\bfx_3\,
    D(\omega,x_{12})\,\psi_a^{\dag}(\bfx_1)\, \alpha_{\mu}
\nonumber \\ & \times
    \bigg[
        G(E,\bfx_1,\bfx_3)\, V_g(\bfx_3)\, G(E,\bfx_3,\bfp)\, \gamma^0 \Sigma^{(0)}_R(E,\bfp)\, G(E,\bfp,\bfx_2)
\nonumber \\ &
 \ \ \
     -
        G^{(0)}(E,\bfx_1,\bfx_3)\, V_g(\bfx_3)\, G^{(0)}(E,\bfx_3,\bfp)\, \gamma^0 \Sigma^{(0)}_R(E,\bfp)\, G^{(0)}(E,\bfp,\bfx_2)
     \bigg]
       \alpha^{\mu} \psi_a(\bfx_2)\,.
\end{align}
In order to isolate the IR divergencies, we subtract and re-add the
following terms in the brackets above,
\begin{align}
    \bigg[\ldots\bigg] = & \ \bigg[\ldots
     \mp
        G^{(a)}(E,\bfx_1,\bfx_3)\, V_g(\bfx_3)\, G^{(a)}(E,\bfx_3,\bfp)\, \gamma^0 \Big[
                    \Sigma^{(0)}_R(\vare_a,\bfp)-\omega\, \Sigma^{(0)^{\prime}}_R(\vare_a,\bfp)\Big]
                                                    \, G^{(a)}(E,\bfp,\bfx_2)
\nonumber \\ &
 \ \ \
     \mp
        G^{(a)}(E,\bfx_1,\bfx_3)\, V_g(\bfx_3)\, G_{{\rm red},a}(\bfx_3,\bfp)\, \gamma^0
                    \Sigma^{(0)}_R(\vare_a,\bfp)
                                                    \, G^{(a)}(E,\bfp,\bfx_2)
                                                    \bigg] \,,
\end{align}
where 
$G_{{\rm red},a}$ is the reduced Green function for the state $a$,
\begin{align}
G_{{\rm red},a}(\bfx_1,\bfx_2) = \sum_{n\neq a} \frac{\psi_n(\bfx_1)\,\psi_n^{\dag}(\bfx_2)}{\vare_a-\vare_n}\,.
\end{align}
The above subtraction renders the main contribution IR finite; the subtracted terms are evaluated separately.
We thus separate $E^P_{\mathrm{NV1}}$ into the sum of the regular (R) and the two infrared
($\mathrm{IR'}$, IR)
parts,
\begin{align}
  E^P_{\mathrm{NV1}} =
  E^P_{\mathrm{NV1}}(\mathrm{R}) + E^P_{\mathrm{NV1}}(\mathrm{IR'}) + E^P_{\mathrm{NV1}}(\mathrm{IR})\,.
\end{align}
The IR contributions are evaluated as
\begin{align}
E^P_{\mathrm{NV1}}(\mathrm{IR'}) = 2\,
\lbr V_g \rbr\,\big<  \gamma^0\Sigma^{(0)}_R(\vare_a)\big>\,
    \bigg[ \frac{\alpha}{4\mu}
    - \frac{\alpha}{4}  \sum_{\mu_{a'}} s_{\mu_{a'}}
     \bra{aa'} \alpha_{\mu}\alpha^{\mu}x_{12}\ket{a'a}
      \bigg]\,,
\end{align}
where the expectation value is to be taken with respect to the reference state, $\langle \dots \rangle = \langle a | \dots | a \rangle$, and
\begin{align}
 s_{\mu_{a'}} = \frac{\bra{a'}V_g\ket{a'}}{\bra{a}V_g\ket{a}} =
  (-1)^{\mu_a-\mu_{a'}}\,C^{10}_{j_a\mu_{a'},j_a-\mu_{a'}}\,\big[C^{10}_{j_a\mu_{a},j_a-\mu_{a}}\big]^{-1}\,.
\end{align}
Furthermore,
\begin{align}
E^P_{\mathrm{NV1}}(\mathrm{IR}) = 2\,
\Big[
 \lbr V_g \rbr\,   \lbr \gamma^0\Sigma^{(0)^{\prime}}_R(\vare_a)\rbr
 + \bra{\delta a} \gamma^0\Sigma^{(0)}_R(\vare_a)\ket{a}
 \Big] \,
  \frac{\alpha}{\pi}
    \bigg[ \ln\frac{\mu}{2}+ \gamma
   +  \sum_{\mu_{a'}} s_{\mu_{a'}}
     \bra{aa'} \alpha_{\mu}\alpha^{\mu}\ln x_{12}\ket{a'a}
      \bigg]\,,
\end{align}
where $\delta a$ is the magnetically perturbed reference-state wave function given by Eq.~(\ref{eq:pwf}).

\subsection{$\bm{E^{P}_{\rm NV2}}$}

The second nested vertex $P$-term contribution is diagrammatically represented
by the second top  line in Fig.~\ref{fig:Pvertex}.
It is given by
\begin{align}
E^P_{\mathrm{NV2}} = & \
    4i\alpha \int_{C_F} d\omega
    \int \frac{d\bfp_1}{(2\pi)^3} \frac{d\bfp_2}{(2\pi)^3} \int d\bfx_1d\bfx_2d\bfx_3\,
    D(\omega,x_{12})\,\psi_a^{\dag}(\bfx_1)\, V_C(q)\,\alpha_{\mu}
\nonumber \\ & \times
    \bigg[
        G(E,\bfx_1,\bfx_3)\, V_g(\bfx_3)\, G(E,\bfx_3,\bfp_1)\, \gamma^0 \Gamma(E,\bfp_1;E,\bfp_2)\, G(E,\bfp_2,\bfx_2)
     \bigg]
       \alpha^{\mu} \psi_a(\bfx_2)\,.
\end{align}
In order to isolate the IR divergencies, we subtract and then re-add the following terms in
the brackets above
\begin{align}
    \bigg[\ldots\bigg] = & \ \bigg[\ldots
     \mp
        G^{(a)}(E,\bfx_1,\bfx_3)\, V_g(\bfx_3)\, G^{(a)}(E,\bfx_3,\bfp_1)\, \gamma^0 \Big[
                    \Gamma(\vare_a,\bfp_1;\vare_a,\bfp_2)
                          -\omega\, \Gamma^{\prime}(\vare_a,\bfp_1;\vare_a,\bfp_2)\Big]
                                                    \, G^{(a)}(E,\bfp_2,\bfx_2)
\nonumber \\ &
 \ \ \
     \mp
        G^{(a)}(E,\bfx_1,\bfx_3)\, V_g(\bfx_3)\, G_{{\rm red},a}(\bfx_3,\bfp_1)\, \gamma^0
                    \Gamma(\vare_a,\bfp_1;\vare_a,\bfp_2)
                                                    \, G^{(a)}(E,\bfp_2,\bfx_2)
 \bigg]\,,
\end{align}
where $\Gamma^{\prime}(\vare_a,\bfp_1;\vare_a,\bfp_2) = d/(d\vare) \Gamma(\vare,\bfp_1;\vare,\bfp_2)|_{\vare = \vare_a}$.
$E^P_{\mathrm{NV2}}$ is thus separated into the regular and two infrared parts,
\begin{align}
E^P_{\mathrm{NV2}} = E^P_{\mathrm{NV2}}(\mathrm{R}) + E^P_{\mathrm{NV2}}(\mathrm{IR'}) + E^P_{\mathrm{NV2}}(\mathrm{IR})\,.
\end{align}
The IR contributions are evaluated as follows
\begin{align}
E^P_{\mathrm{NV2}}(\mathrm{IR'}) = 2\,
\lbr V_g \rbr\,\big<  \gamma^0\,V_C\,\Gamma(\vare_a;\vare_a)\big>\,
    \bigg[ \frac{\alpha}{4\mu}
    - \frac{\alpha}{4}  \sum_{\mu_{a'}} s_{\mu_{a'}}
     \bra{aa'} \alpha_{\mu}\alpha^{\mu}x_{12}\ket{a'a}
      \bigg]\,,
\end{align}
and
\begin{align}
E^P_{\mathrm{NV2}}(\mathrm{IR}) = 2\,
\Big[
 \lbr V_g \rbr\,   \lbr \gamma^0\,V_C\,\Gamma^{\prime}(\vare_a;\vare_a)\rbr
 + \bra{\delta a} \gamma^0\,V_C\,\Gamma(\vare_a;\vare_a)\ket{a}
 \Big] \,
  \frac{\alpha}{\pi}
    \bigg[ \ln\frac{\mu}{2}+ \gamma
   +  \sum_{\mu_{a'}} s_{\mu_{a'}}
     \bra{aa'} \alpha_{\mu}\alpha^{\mu}\ln x_{12}\ket{a'a}
      \bigg]\,.
\end{align}
The regular contribution is finite and can be evaluated numerically as it is. However, it is convenient
to separate it into two parts, the one diagonal in the momentum (with $\vec{p}_1 = \vec{p}_2$)
and the nondiagonal one, according to Eq.~(\ref{eq:vertex_separation}).
In the diagonal part the integration over one momentum is
factorized out. It is transformed to the following form, favourable for
numerical evaluation,
\begin{align}\label{eq:NV2}
E^P_{\mathrm{NV2}}(\mathrm{diag}) =
    -4i\alpha \int_{C_F} d\omega
     & \,
    \int \frac{d\bfp}{(2\pi)^3} \int d\bfx_1d\bfx_2d\bfx_3\,
    D(\omega,x_{12})\,\psi_a^{\dag}(\bfx_1)\, \alpha_{\mu}
\nonumber \\ & \times
        G(E,\bfx_1,\bfx_3)\, V_g(\bfx_3)\,
           G(E,\bfx_3,\bfp)\, \gamma^0\, \Sigma^{(0)^{\prime}}_R(E,\bfp)\, G_V(E,\bfp,\bfx_2)\,
       \alpha^{\mu} \psi_a(\bfx_2)\,.
\end{align}
We note that in the above expression
the would-be Coulomb singularity of the nuclear potential $V_C$ is absorbed in $G_V$,
where it does not cause any computational difficulties.

\subsection{$\bm{E^{P}_{\rm NV3}}$}
The third nested vertex $P$-term contribution is diagrammatically represented
by the third line in Fig.~\ref{fig:Pvertex}.
Its expression is
\begin{align} \label{eq:nv3}
E^P_{\mathrm{NV3}} = & \
    2i\alpha \int_{C_F} d\omega
    \int \frac{d\bfp_1}{(2\pi)^3} \frac{d\bfp_2}{(2\pi)^3} \int d\bfx_1d\bfx_2\,
    D(\omega,x_{12})\,\psi_a^{\dag}(\bfx_1)\, \alpha_{\mu}
\nonumber \\ & \times
    \bigg[
        G(E,\bfx_1,\bfp_1)\, \gamma^0 \Lambda_{\rm Zee}(E,\bfp_1;E,\bfp_2)\, G(E,\bfp_2,\bfx_2)
-  G^{(0)}(E,\bfx_1,\bfp_1)\, \gamma^0 \Lambda_{\rm Zee}(E,\bfp_1;E,\bfp_2)\, G^{(0)}(E,\bfp_2,\bfx_2)
     \bigg]
       \alpha^{\mu} \psi_a(\bfx_2)
       \,,
\end{align}
where $\Lambda_{\rm Zee}(\vare_1,\bfp_1;\vare_2,\bfp_2)$ is the free vertex function with the
Zeeman magnetic interaction
(which will be referred to as the Zeeman vertex); its explicit expression will be given below in
Eq.~(\ref{eq:lambda_zee}).
We separate out the infrared divergences by subtracting and re-adding the following contribution in the
brackets,
\begin{align}
    \bigg[\ldots\bigg] = & \ \bigg[\ldots
     \mp
        G^{(a)}(E,\bfx_1,\bfp_1)\, \gamma^0
                    \Lambda_{\rm Zee}(\vare_a,\bfp_1;\vare_a,\bfp_2)
                    \, G^{(a)}(E,\bfp_2,\bfx_2)
                    \bigg]\,,
\end{align}
which separates Eq.~(\ref{eq:nv3}) into two parts,
\begin{align}
E^P_{\mathrm{NV3}} =
        E^P_{\mathrm{NV3}}(\mathrm{R}) + E^P_{\mathrm{NV3}}(\mathrm{IR})\,.
\end{align}
The IR contribution is evaluated as
\begin{align}
E^P_{\mathrm{NV3}}(\mathrm{IR}) =
\lbr \gamma^0\,\Lambda_{\rm Zee}(\vare_a;\vare_a)\rbr\,
  \frac{\alpha}{\pi}
    \bigg[ \ln\frac{\mu}{2}+ \gamma
   +  \sum_{\mu_{a'}} s_{\mu_{a'}}
     \bra{aa'} \alpha_{\mu}\alpha^{\mu}\ln x_{12}\ket{a'a}
      \bigg]\,.
\end{align}

We now turn to the Zeeman vertex function $\Lambda_{\rm Zee}$.
The straightforward representation of its matrix element
with the Dirac wave functions is given by Eq.~(15) of
Ref.~\cite{yerokhin:04}. This
representation involves the gradient of the Dirac $\delta$ function and is cumbersome to
work with. In the present work we choose to follow Ref.~\cite{blundell:97:pra} and to use the
regularized version of the Zeeman magnetic interaction. Specifically, we introduce a small regulator parameter $\rho$
and define the regularized magnetic potential $V_{g,\rho}$ in the coordinate space as
\begin{align}
V_{g,\rho}(\bfr) = \frac{1}{\mu_a}\,[\bfr \times \balpha]_z\, e^{-(\rho r/2)^2}\,,
\end{align}
which obviously approaches $V_g$ as $\rho \to 0$. Performing the Fourier transform,
we obtain the matrix element of $V_{g,\rho}$
with the Dirac wave functions in momentum space as
\begin{align}
\bra{n} V_{g,\rho} \ket{m} =&\, \int d\bfx\, \psi_n^{\dag}(\bfx)\, V_{g,\rho}(\bfx)\, \psi_m(\bfx)
  = (-4\pi i)\,A_{\rho}
   \int \frac{d\bfp_1}{(2\pi)^3} \frac{d\bfp_2}{(2\pi)^3} \,
    \psi_n^{\dag}(\bfp_1)\, \gamma^0\,
    [\bfq \times \bgamma]_z\,
     e^{-q^2/\rho^2}\, \psi_m(\bfp_2)\,,
\end{align}
where $\bfq = \bfp_1-\bfp_2$, $q = |\bfq|$, and  $A_{\rho} = 4\pi^{1/2}/(\mu_a\,\rho^5)$.
Analogously, the matrix element of the Zeeman vertex between the Dirac states in momentum space is given by
\begin{align} \label{eq:lambda_zee}
\bra{n} \gamma^0 \Lambda_{\rm Zee}(\vare_1,\vare_2) \ket{m} =&\,
   (-4\pi i)\,A_{\rho}
   \int \frac{d\bfp_1}{(2\pi)^3} \frac{d\bfp_2}{(2\pi)^3} \,
    \psi_n^{\dag}(\bfp_1)\, \gamma^0\,
    [\bfq \times \vec{\Gamma}(\vare_1,\bfp_1;\vare_2,\bfp_2)]_z\,
     e^{-q^2/\rho^2}\, \psi_m(\bfp_2)\,,
\end{align}
where $\vec{\Gamma}(\vare_1,\bfp_1;\vare_2,\bfp_2)$ is the vector part of the renormalized
one-loop vertex function defined
in Appendix A of Ref.~\cite{yerokhin:99:sescr}.
Note that the regularized Zeeman vertex has essentially the same form as the vertex with the magnetic
dipole hyperfine interaction, see Eq.~(28) of Ref.~\cite{yerokhin:10:sehfs}.
The angular integrations of expressions with
the Zeeman vertex in momentum space are detailed out in Appendix \ref{app:verzee}.

In actual numerical calculations, we found that when reasonably small values of the regulator
($\rho \approx 10^{-4}$) are used, the dependence of results on the regulator value is negligible.
In addition, the integration over one of the two momenta (say, $\bfp_2$)
is greatly simplified by the fact that the integrand is strongly peaked
around the point $\bfp_2 = \bfp_1$, so that the integration
is actually needed only over a small region $\bfp_2 \approx
\bfp_1$.

\subsection{$\bm{E^{P}_{\rm OV}}$}

The overlapping vertex $P$-term contribution is
represented by the bottom line of Fig.~\ref{fig:Pvertex} and
is given by
\begin{align} \label{eq:POV}
E^P_{\mathrm{OV}} = & \
    -4i\alpha \int_{C_F} d\omega
    \int \frac{d\bfp_1}{(2\pi)^3} \frac{d\bfp_2}{(2\pi)^3} \int d\bfz\, d\bfy\,
    \frac{e^{-i\bfq\cdot\bfz}}{\omega^2-\bfq\,^2 + i0}\,
    \psi_a^{\dag}(\bfz)\, \alpha_{\mu}
\nonumber \\ & \times
    \Big[
        G(E,\bfz,\bfy)\, V_g(\bfy)\,G(E,\bfy,\bfp_1)
    - G^{(0)}(E,\bfz,\bfy)\, V_g(\bfy)\,G^{(0)}(E,\bfy,\bfp_1)
    \Big]
         \gamma^0 \Gamma^{\mu}_R(E,\bfp_1;\vare_a,\bfp_2)\, \psi_a(\bfp_2)
         \,.
\end{align}
In order to separate out the IR divergency, we subtract and re-add the following contribution in the
brackets
\begin{align} \label{eq:POVinfr}
\Big[ \ldots \Big] = \Big[ \ldots  \mp
   G^{(a)}(E,\bfz,\bfy)\, V_g(\bfy)\,G^{(a)}(E,\bfy,\bfp_1) \Big]
\,,
\end{align}
thus separating $E^P_{\mathrm{OV}}$ into the regular (R) and infrared (IR) parts,
\begin{align}
E^P_{\mathrm{OV}} = E^P_{\mathrm{OV}}(\mathrm{R}) + E^P_{\mathrm{OV}}(\mathrm{IR})\,.
\end{align}
In the regular contribution, we perform the Wick rotation of the $\omega$ integration contour,
separating pole contributions as follows (assuming that $a$ is the
ground state)
\begin{align}
i \sum_{n_1n_2 \neq aa}\int_{C_F} d\omega\,
\frac{f_{n_1n_2}(\omega)}{(\vare_a-\omega-u\vare_{n_1})(\vare_a-\omega-u\vare_{n_2})}
 = -2\, \bigg\{ &\,
  -\frac{\pi}{2}\sum_{n_2\neq a}\frac{f_{an_2}(0)}{\Delta_{an_2}}
 -\frac{\pi}{2}\sum_{n_1\neq a}\frac{f_{n_1a}(0)}{\Delta_{an_1}}
\nonumber \\ &
 + {\rm \Re}\, \sum_{n_1n_2 \neq aa} \int_0^{\infty}d\omega\,
 \frac{f_{n_1n_2}(i\omega)}{(\Delta_{an_1}-i\omega)(\Delta_{an_2}-i\omega)}
 \bigg\}\,,
\end{align}
where $\Delta_{an} = \vare_a - \vare_n$ and in the summation over $n_1$ and $n_2$ the term with $(n_1,n_2) = (a,a)$
is omitted.
The angular integration in $E^P_{\mathrm{OV}}$ is quite complicated and is detailed out in Appendix~\ref{app:POV}.
The IR contribution is transformed as
\begin{align}
E^P_{\mathrm{OV}}({\mathrm{IR}}) = & \
    -4i\alpha \int_{C_F} d\omega
    \frac1{(-\omega + i0)^2}
    \int \frac{d\bfp_1}{(2\pi)^3} \frac{d\bfp_2}{(2\pi)^3}
    \sum_{\mu_{a'}}
    \bra{a'} V_g \ket{a'}\,
\nonumber \\ & \times
    \bigg(
     \int d\bfz
    \frac{e^{-i\bfq\cdot\bfz}}{\omega^2-\bfq\,^2 + i0}\,
    \psi_a^{\dag}(\bfz)\, \alpha_{\mu} \,\psi_{a'}(\bfz)\,
     \bigg)\,
        \psi_{a'}^{\dag}(\bfp_1)\,
         \gamma^0 \Gamma^{\mu}_R(E,\bfp_1;\vare_a,\bfp_2)\, \psi_a(\bfp_2)
         \,,
\end{align}
where the state $a'$ differs from the reference state $a$ only by the angular-momentum projection $\mu_{a'}$.

The IR divergence present in the above expression cancels out when $E^P_{\mathrm{OV}}({\mathrm{IR}})$ is considered together
with the corresponding contribution from the overlapping derivative term, $E^P_{\mathrm{OD}}({\mathrm{IR}})$.
We, therefore, introduce the total overlapping IR contribution as
\begin{align}
E^P_{\mathrm{OVD}}({\mathrm{IR}}) = E^P_{\mathrm{OV}}({\mathrm{IR}}) + E^P_{\mathrm{OD}}({\mathrm{IR}})
\,.
\end{align}
In order to calculate $E^P_{\mathrm{OVD}}({\mathrm{IR}})$ numerically, we need to deform the $\omega$ integration
contour to be parallel to the imaginary axis. However, the standard Wick rotation is not possible in this
case. The situation is somewhat similar to that for the vertex $+$ reducible contribution for the
one-loop self-energy correction to the $g$ factor. In Ref.~\cite{blundell:97:pra}, it was suggested
to use a small numerical regularization parameter to shift the reference-state energy, thus removing
the infrared divergence of the integral. The idea was to perform the Wick rotation of the regularized
expression and then evaluate the limit numerically by decreasing the regularization parameter.
In the present work, we use a similar but somewhat different approach. We introduce
a small regularization parameter $\epsilon$ that shifts the Wick-rotated $\omega$ integration contour,
$(-i\infty,i\infty) \to (\epsilon-i\infty,\epsilon+i\infty)$.
We obtain the regularized expression
\begin{align} \label{POV:IR}
E^P_{\mathrm{OVD}}({\mathrm{IR}}, \epsilon) = & \
    8\alpha\, \mathrm{Re}\,\int_0^{\infty} dw\,
    \frac1{\omega^2}
    \int \frac{d\bfp_1}{(2\pi)^3} \frac{d\bfp_2}{(2\pi)^3}
    \Big[\sum_{\mu_{a'}}\bra{a'} V_g \ket{a'}\, - \delta_{\mu_{a}\mu_{a'}} \bra{a} V_g \ket{a}\Big]
\nonumber \\ & \times
    \bigg(
     \int d\bfz
    \frac{e^{-i\bfq\cdot\bfz}}{\omega^2-\bfq\,^2}\,
    \psi_a^{\dag}(\bfz)\, \alpha_{\mu} \,\psi_{a'}(\bfz)\,
     \bigg)\,
        \psi_{a'}^{\dag}(\bfp_1)\,
         \gamma^0 \Gamma^{\mu}_R(\vare_a-\omega,\bfp_1;\vare_a,\bfp_2)\, \psi_a(\bfp_2)
         \Bigg|_{\omega = \epsilon + iw}
         \,.
\end{align}
Here, the integrand is supposed to be evaluated for $\omega = \epsilon + iw$. The final infrared
contribution is determined by numerically evaluating the limit $\epsilon \to 0$,
\begin{align} \label{POV:IR2}
E^P_{\mathrm{OVD}}({\mathrm{IR}}) = \lim_{\epsilon \to 0+} E^P_{\mathrm{OVD}}({\mathrm{IR}}, \epsilon)\,.
\end{align}
Specifically, we perform calculations for several
values of the regulator parameter (typically, about 10 values in the range between $10^{-3}$ and
$10^{-2}$) and obtain the final IR contribution by a polynomial fit of the
limit $\epsilon \to 0$.

It is important that the limit in Eq.~(\ref{POV:IR2}) is approached from the positive values of the
regulator. It is possible to use the negative values of the regulator as well, but then one
needs to add the pole contribution at $\omega = 0$, according to the identity
\begin{align}
 -2\, \lim_{\epsilon \to 0+} \mathrm{Re} \int_0^{\infty} dw\, \frac{f(\epsilon+ iw)}{(\epsilon +iw)^2}
=
 -2\,  \lim_{\epsilon \to 0-} \mathrm{Re} \int_0^{\infty} dw\, \frac{f(\epsilon+ iw)}{(\epsilon +iw)^2}
-2 \pi\, f^{\prime}(0)\,.
\end{align}
We checked the fulfillment of this identity numerically, which served as an additional check of our
numerical procedures. It
is  interesting that for the one-loop self-energy correction to the $g$ factor, the pole contribution
vanishes, so that the limit $\epsilon \to 0$ can be approached from both the positive and the negative
sides, with the same result.

We also mention that the above contour with $\epsilon > 0$ can be considered as a variant of our
standard contour $C_{\rm LH}$, in which the length of the low-energy part of the contour (and, therefore,
the corresponding low-energy contribution) approaches zero.

%

\section{Additional contribution}

The additional $P$-term contribution is introduced in this work in order to make the whole $P$ term IR finite.
We will demonstrate that the IR divergences in individual $P$-term contributions discussed in the previous sections
disappear when combined with (parts of) contributions referred to as the loop-after-loop reducible
part in Ref.~\cite{sikora:18:phd} and given by Eqs.~(6.17)-(6.20) of that work.
Specifically, we define the additional $P$-term contribution as
\begin{align}
E^{P}_{\rm ADD} = &\, \Big\{
 \lbr V_g \rbr\,\big<  \gamma^0\Sigma^{(0)^{\prime}}_R(\vare_a)\big>
  + \lbr \gamma^0\,\Lambda_{\rm Zee}(\vare_a;\vare_a)\rbr
  + 2 \bra{\delta a} \gamma^0 \big[ \Sigma^{(0)}_R(\vare_a)
  + V_C\,\Gamma(\vare_a;\vare_a)\big] \ket{a}
 \Big\}
 \lbr \gamma^0 \Big[
 \Sigma^{\prime}(\vare_a) -  \Sigma^{(0)^{\prime}}_R(\vare_a)
 \Big] \rbr\,.
\end{align}

We split $E^{P}_{\rm ADD}$ into the regular and IR parts,
$E^{P}_{\rm ADD} = E^{P}_{\rm ADD}(\mathrm{R}) + E^{P}_{\rm ADD}(\mathrm{IR})$,
by subtracting and re-adding in the matrix element of the self-energy operator $\Sigma(\vare)$
the contribution of the reference states $\Sigma^{(a)}(\vare)$,
\begin{align}
 \lbr \gamma^0 \Big[
 \Sigma^{\prime}(\vare_a) -  \Sigma^{(0)^{\prime}}_R(\vare_a)
 \Big] \rbr
=
 \lbr \gamma^0 \Big[
 \Sigma^{\prime}(\vare_a) -  \Sigma^{(0)^{\prime}}_R(\vare_a)
 -  \Sigma^{(a)^{\prime}}(\vare_a)
 \Big] \rbr
+
 \lbr \gamma^0 \Sigma^{(a)^{\prime}}(\vare_a) \rbr\,.
\end{align}
Here,
\begin{align}
\lbr \gamma^0 \Sigma^{(a)}(\vare)\rbr = \frac{i}{2\pi}\int_{-\infty}^{\infty} d\omega  \sum_{\mu_{a'}}
 \frac{\bra{aa'} I(\omega) \ket{a'a}}{\vare-\omega-\vare_a+i0}\,.
\end{align}
Using Eq.~(\ref{eq:Ja}), we immediately evaluate the infrared contribution as
\begin{align}
E^{P}_{\rm ADD}(\mathrm{IR}) =  - \Big\{
 \lbr V_g \rbr\,\big<  \gamma^0\Sigma^{(0)^{\prime}}_R(\vare_a)\big>
  + \lbr \gamma^0\,\Lambda_{\rm Zee}(\vare_a;\vare_a)\rbr
  + 2 \bra{\delta a} \gamma^0 \big[ \Sigma^{(0)}_R(\vare_a)
  + V_C\,\Gamma(\vare_a;\vare_a)\big] \ket{a}
 \Big\} \, J_2\,.
\end{align}
The regular part $E^{P}_{\rm ADD}(\mathrm{R})$ is calculated to a very
high accuracy by a straightforward generalization of the method
developed for the one-loop self-energy in Ref.~\cite{yerokhin:05:se}.

\end{widetext}

%

\section{Total $\bm{P}$ term}

Finally, the total $P$-term is given by the sum of all contributions,
\begin{align}
E_{P} =&\ E^{P}_{\rm NW1} + E^{P}_{\rm NW2} + E^{P}_{\rm OW} + E^{P}_{\rm ND1} + E^{P}_{\rm ND2} + E^{P}_{\rm ND3}
\nonumber \\ &
+ E^{P}_{\rm NV1} + E^{P}_{\rm NV2} + E^{P}_{\rm NV3}
+ E^{P}_{\rm OD} + E^{P}_{\rm OV}
+ E^{P}_{\rm ADD} \,.
\end{align}
The $P$ term $E_{P}$ is finite. The IR divergence in individual contributions
disappears when all terms are added together.
The cancelation of divergences between the individual contributions is
demonstrated in Table~\ref{tab:div}.

\begin{table*}
\caption{Cancelation of infrared-divergent parts of individual $P$-term contributions.
\label{tab:div}}
\begin{ruledtabular}
\begin{tabular}{lcc}
\multicolumn{1}{l}{Contribution} &
        \multicolumn{1}{c}{$\times \frac{\alpha}{4\mu}$} &
            \multicolumn{1}{c}{$ \times \frac{\alpha}{\pi}\big( \ln \frac{\mu}{2} + \gamma \big)$}
\\
\hline\\[-5pt]
$E^P_{\mathrm{ND1}}$  & $(-2)\,\lbr V_g \rbr\,\big<  \gamma^0\Sigma^{(0)}_R(\vare_a)\big>$
                      & $(-2)\,\lbr V_g \rbr\,\big<  \gamma^0\Sigma^{(0)^{\prime}}_R(\vare_a)\big> $\\[5pt]
$E^P_{\mathrm{ND2}}$  & $(-2)\,\lbr V_g \rbr\,\big< \gamma^0\,V_C\, \Gamma(\vare_a;\vare_a)\big>$
                      & $(-2)\,\lbr V_g \rbr\,\big< \gamma^0\,V_C\, \Gamma^{\prime}(\vare_a;\vare_a)\big>$\\[5pt]
$E^P_{\mathrm{ND3}}$  &
                      & $\lbr V_g \rbr\,\big<  \gamma^0\Sigma^{(0)^{\prime}}_R(\vare_a)\big> $\\[5pt]
$E^P_{\mathrm{NV1}}$  & $ 2\,\lbr V_g \rbr\,\big<  \gamma^0\Sigma^{(0)}_R(\vare_a)\big>$
                      & $ 2\,\lbr V_g \rbr\,\big<  \gamma^0\Sigma^{(0)^{\prime}}_R(\vare_a)\big>
                        + 2 \bra{\delta a} \gamma^0\Sigma^{(0)}_R(\vare_a)\ket{a}$\\[5pt]
$E^P_{\mathrm{NV2}}$
                      & $2\,\lbr V_g \rbr\,\big<  \gamma^0\,V_C\,\Gamma(\vare_a;\vare_a)\big>$
                      & $2\, \lbr V_g \rbr\,   \lbr \gamma^0\,V_C\,\Gamma^{\prime}(\vare_a;\vare_a)\rbr
                        + 2\bra{\delta a} \gamma^0\,V_C\,\Gamma(\vare_a;\vare_a)\ket{a}$ \\[5pt]
$E^P_{\mathrm{NV3}}$
                      &
                      & $\lbr \gamma^0\,\Lambda_{\rm Zee}(\vare_a;\vare_a)\rbr$
                      \\[5pt]
$E^P_{\mathrm{ADD}}$
                      & &
 $-\lbr V_g \rbr\,\big<  \gamma^0\Sigma^{(0)^{\prime}}_R(\vare_a)\big>
 - \lbr \gamma^0\,\Lambda_{\rm Zee}(\vare_a;\vare_a)\rbr
  - 2 \bra{\delta a} \gamma^0\big[ \Sigma^{(0)}_R(\vare_a)
  +V_C\,\Gamma(\vare_a;\vare_a)\big] \ket{a}$
  \\[5pt]
                      \hline\\[-5pt]
Sum                   &   0
 & 0
\end{tabular}
\end{ruledtabular}
\end{table*}

\section{Numerical calculations}

The $P$ term consists of many different contributions whose computation is complicated and time-consuming.
In order to obtain reliable numerical results, we had to perform numerous cross-checks of our
numerical procedures and test calculations, of which some have already been mentioned.
Others will be discussed in this section.

We performed a set of test calculations, in which
we reduced our two-loop $P$-term contributions to
less complicated one-loop diagrams.
More specifically,
we observed that in each of the $P$-term contributions, if we ``collapse'' the
inner one-loop subgraph and replace it by the
corresponding vertex, the result can be converted back to the coordinate
space and computed by standard methods developed for calculations of one-loop
self-energy
corrections. Specifically, the replacements are $\Sigma^{(0)}_R(\vare,\bfp)\to \gamma^0$,
$\Gamma^{\mu}_R(\vare_1,\bfp_1;\vare_2,\bfp_2) \to \gamma^{\mu}$,
$ \Lambda_{\rm Zee}(\vare_1,\bfp_1;\vare_2,\bfp_2) \to \gamma^0\,V_g$.
These simplifications give a very valuable opportunity to check our numerical procedures
for the two-loop $P$-term contributions by reducing them to one-loop calculations
which can be independently checked. We did these tests for all $P$-term contributions
discussed in the previous sections.

\begin{widetext}
As an illustration, let us now examine a test calculation carried out for the most complicated
$P$-term contribution, specifically, the vertex overlapping term
 $E^P_{\mathrm{OV}}$. After performing the above mentioned replacement
$\Gamma^{\mu}_R(\vare_1,\bfp_1;\vare_2,\bfp_2) \to \gamma^{\mu}$, the regular part of
$E^P_{\mathrm{OV}}$ becomes $E^P_{\mathrm{OV}}(\mathrm{R}) \to  \delta E_1
$, where
\begin{align} \label{eq:deltaE1}
& \, \delta E_1
=    -4i\alpha \int_{C_F} d\omega
    \int \frac{d\bfp_1}{(2\pi)^3} \frac{d\bfp_2}{(2\pi)^3} \int d\bfz\, d\bfy\,
    \frac{e^{-i\bfq\cdot\bfz}}{\omega^2-\bfq\,^2 + i0}\,
    \psi_a^{\dag}(\bfz)\, \alpha_{\mu}
\nonumber \\ \times & \,
    \Big[
        G(E,\bfz,\bfy)\, V_g(\bfy)\,G(E,\bfy,\bfp_1)
    - G^{(0)}(E,\bfz,\bfy)\, V_g(\bfy)\,G^{(0)}(E,\bfy,\bfp_1)
    - G^{(a)}(E,\bfz,\bfy)\, V_g(\bfy)\,G^{(a)}(E,\bfy,\bfp_1)
    \Big]
         \alpha^{\mu}\, \psi_a(\bfp_2)
         \,.
\end{align}
On one hand, $\delta E_1$ can be calculated according to Eq.~(\ref{eq:deltaE1})
by the numerical procedure developed for
$E^P_{\mathrm{OV}}(\mathrm{R})$. One the other hand, the above expression can be
transformed to the fully coordinate representation, with the result
\begin{align}
\delta E_1 \label{eq:deltaE1b}
&\,
=    -4i\alpha \int_{C_F} d\omega
\int d\bfz\, d\bfy\,d\bfx\, D(\omega,|\bfz-\bfx|)\,
    \psi_a^{\dag}(\bfz)\, \alpha_{\mu}
\nonumber \\ \times & \,
    \Big[
        G(E,\bfz,\bfy)\, V_g(\bfy)\,G(E,\bfy,\bfx)
    - G^{(0)}(E,\bfz,\bfy)\, V_g(\bfy)\,G^{(0)}(E,\bfy,\bfx)
    - G^{(a)}(E,\bfz,\bfy)\, V_g(\bfy)\,G^{(a)}(E,\bfy,\bfx)
    \Big]
         \alpha^{\mu}\, \psi_a(\bfx)
         \,.
\end{align}
\end{widetext}
Eq.~(\ref{eq:deltaE1b}) can be identified as the (1+)-potential vertex contribution to the
self-energy correction to the $g$ factor, which was encountered in our previous calculations
\cite{yerokhin:02:prl,yerokhin:04}. Computation of this expression in coordinate space is
relatively straightforward and was performed by numerical routines developed in Ref.~\cite{yerokhin:04}.
Numerical results of our test calculations of $\delta E_1$
carried out in the mixed representation
and in the coordinate representation are presented in Table~\ref{tab:comp}.
We observe that our computations in the mixed representation
reproduce well the (much more accurate)
values obtained in the coordinate representation.
This comparison proves that
we are able to keep the momentum integrations in the mixed momentum-coordinate representation
well under control.
It should be pointed out that the two computations used different integration contours
for the $\omega$ integration. In the mixed representation, we used the integration over
the imaginary axis, whereas in the coordinate representation, we used our standard $C_{\rm LH}$
integration contour
\cite{yerokhin:20:green}. So, this test also checks the correctness of our treatment of the pole
contributions in the mixed representation.

\begin{table}
\caption{Results of the test calculation: $\delta E_1 \times \alpha/(4\pi) $
computed in the mixed momentum-coordinate representation,
Eq.~(\ref{eq:deltaE1}), and in the fully coordinate representation, Eq.~(\ref{eq:deltaE1b}).
Units are $10^{-6}$.
\label{tab:comp}}
\begin{ruledtabular}
\begin{tabular}{ldd}
 \multicolumn{1}{l}{$|\kappa|$}
& \multicolumn{1}{c}{Mixed} &
        \multicolumn{1}{c}{Coordinate}
\\
\hline\\[-5pt]
   1             &      1.63831         &   1.63827 \\
   2             &     -0.03713         &  -0.03713 \\
   3             &     -0.04940         &  -0.04940 \\
   4             &     -0.02442         &  -0.02442 \\
   5             &     -0.01327         &  -0.01327 \\
6$\ldots$15
          &            -0.02509         &  -0.02511 \\
16$\ldots\infty$
          &            -0.00377\,(23)   &  -0.00382\,(2) \\
Sum       &             1.48518\,(23)   &   1.48513\,(2) \\
\end{tabular}
\end{ruledtabular}
\end{table}

Another important test of our numerical procedures
was the check of cancelation of the $J = 0$ multipole part of
infrared-divergent contributions.
More specifically,
the IR$'$ and IR contributions contain an expansion over the multipoles $J\ge 0$
of the photon propagator. It was shown \cite{sikora:18:phd} that only
the lowest multipole $J = 0$ part is diverging,
whereas the higher-order multipoles $J > 0$ are actually finite.
(For the $1s$ reference state considered here, there is only one other multipole contribution, $J = 1$.)
It was also shown \cite{sikora:18:phd} that the $J = 0$ multipole contributions cancel
{\em exactly} in the total sum, both
the divergent parts and finite remainders.
In the present work we checked these statements numerically.

Our numerical results obtained for individual $P$-term contributions
for the $1s$ state of hydrogen-like tin ($Z = 50$) and bismuth ($Z = 83$)
are presented in
Table~\ref{tab:fin:50} for $Z = 50$ and Table~\ref{tab:fin:83} for $Z = 83$.
The tables present numerical results for the IR$'$ and IR
contributions in two variants (the left and right subcolumns). The left
subcolumns (labeled ``$J\ge 0$'') show results with the $J=0$ multipole
included, whereas the right subcolumns (labeled ``$J > 0$'')
presents results with the $J=0$ multipole excluded.
We observe that sums of both subcolumns are the same, both for the IR and the IR$'$
contribution (although the $J\ge 0$ case leads to much larger numerical cancelations).
This proves that the $J = 0$ multipole terms cancel each other identically, as they
should.

Furthermore, we used the fact that the $J = 1$ multipole contributions
of the IR$'$ and IR terms are finite for an additional test.
We performed calculations of the IR$'$ and IR terms for the nested diagrams
in two different ways.
First, we computed the $J = 1$ contributions by the formulas
presented above, with the $\omega$ integration calculated analytically.
Second, we
computed the same contributions by the general routine developed for the regular parts, with the
$\omega$ integration performed numerically.
Perfect agreement was observed in all cases.
As a matter of fact, the explicit separation of the IR contributions was not
even necessary. We could have removed all IR subtractions and
just dropped the $J = 0$ multipole in our
general codes computing the regular contributions, and the
total result would be the same.
However, we found it safer
to compute the IR contributions separately and check their cancelation explicitly.

For the overlapping-diagram IR contributions, we were not able to carry out the analytical calculations
of the $\omega$ integrals
and had to rely on the numerical integration method. In order to cross-check this part,
we used numerical integration with both positive and negative values of the
regulator $\epsilon$, as discussed in the previous section.

The numerical approach used in this work for computing Feynman diagrams in the mixed
momentum-coordinate representation closely follows the
procedure described in Ref.~\cite{yerokhin:10:sese} and is therefore not
discussed further here.

%
%
\section{Results}

Tables~\ref{tab:fin:50} and \ref{tab:fin:83} summarize our
numerical results obtained for individual $P$-term contributions
for the $1s$ state of hydrogen-like tin ($Z = 50$) and bismuth ($Z = 83$).
Our calculations were performed for the point nuclear model.
Each $P$-term contribution is separated into the infrared IR$'$, IR
and the regular R parts.
The infrared contributions are calculated in two variants, with all multipoles
of the photon propagator included (``$J\ge 0$'' columns) and
with the $J = 0$ multipole excluded (``$J > 0$'' columns), see
the discussion in the previous section. The sum of all individual
IR$'$ and IR contributions in the $J\ge 0$ and $J > 0$ columns
is the same within the numerical uncertainty, which served as an
important cross-check of our calculations.

We note large numerical cancelation between individual $P$-term
contributions, particularly pronounced for smaller values of $Z$.
For instance, for $Z = 50$, the sum of absolute values of all $P$-term contributions
is by about three orders of magnitude larger than the final result.
This cancelation highlights the necessity of carefully controlling numerical
uncertainties at intermediate stages of the computation.
Any small mistake in a single contribution would have
a greatly magnified impact on the final result.

The dominant numerical uncertainty in our computations
comes from the overlapping vertex contribution $E^{P}_{\rm OV}$.
The limitation was due to the fact
that we were able to extend the partial-wave summation
only up to $|\kappa_{\max}| = 15$. For larger values of $\kappa$,
the computation of the angular-momentum coefficients described in Appendix~\ref{app:POV}
becomes unstable due to severe numerical cancelations
arising for $q\to 0$.
Even with extended-precision arithmetic for the angular-momentum coefficients,
we were unable to extend the partial-wave summation beyond this point.

\begin{table*}
\caption{Numerical results for the finite parts of individual $P$-term contributions, for
$Z = 50$,
in units of $10^{-6}$.
\label{tab:fin:50}}
\begin{ruledtabular}
\begin{tabular}{lw{4.6}w{4.6}w{4.6}w{4.6}w{4.8}w{4.8}}
\multicolumn{1}{l}{Term} &
        \multicolumn{2}{c}{$\mathrm{IR'}$} &
            \multicolumn{2}{c}{$\mathrm{IR}$} &
                \multicolumn{1}{c}{R} &
                \multicolumn{1}{c}{Sum}
                \\[3pt]
\cline{2-3}
\cline{4-5}

\\[-7pt]
 & \multicolumn{1}{c}{$J \ge 0$}
 & \multicolumn{1}{c}{$J > 0$}
 & \multicolumn{1}{c}{$J \ge 0$}
 & \multicolumn{1}{c}{$J > 0$}
 &
 &
\\ \hline\\[-5pt]
$\mathrm{NW1}$  &  \cent{-}    &  \cent{-}    &  \cent{-} &  \cent{-} &     1.8363\,(2) &     1.8363\,(2) \\
$\mathrm{NW2}$  &  \cent{-}    &  \cent{-}    &  \cent{-} &  \cent{-} &    -1.2483\,(1) &    -1.2483\,(1) \\
$\mathrm{OW}$   &  \cent{-}    &  \cent{-}    &  \cent{-} &  \cent{-} &     0.5624      &     0.5624     \\
$\mathrm{ND1}$  & -21.1790     & -0.7672      & -36.3756  &  -0.9667  &   -25.0694\,(6) &   -26.8033\,(6)\\
$\mathrm{ND2}$  &  14.0526     &  0.5090      & -19.7867  &  -0.5258  &   -12.0327\,(4) &   -12.0495\,(4)\\
$\mathrm{ND3}$  &  \cent{-}    &  \cent{-}    &  18.1878  &   0.4833  &    14.5088      &    14.9921  \\
$\mathrm{NV1}$  &  20.1561     & -0.2557      &  38.0456  &  -0.3494  &    22.5079\,(6) &    21.9028\,(6)\\
$\mathrm{NV2}$  & -13.3739     &  0.1697      &  18.9901  &  -0.1744   &   12.5088\,(4) &    12.5041\,(4) \\
$\mathrm{NV3}$  &  \cent{-}    &  \cent{-}    & -11.7391  &   0.1078  &   -11.6888\,(4) &   -11.5810\,(4) \\
$\mathrm{OD}+\mathrm{OV}$  &  \cent{-}  &  \cent{-}
                                              &   0.3602  &   0.3602  &     1.7801\,(40)&     2.1403\,(40) \\
$\mathrm{ADD}$  &  \cent{-}    &  \cent{-}    &  -8.9860  &  -0.2388  &    -2.2846      &    -2.5234\\
\hline\\[-5pt]
Sum             & -0.3442      &   -0.3442    &  -1.3037  &  -1.3038  &    1.3805\,(42) &     -0.2674\,(42) \\
\end{tabular}
\end{ruledtabular}
\end{table*}

\begin{table*}
\caption{Numerical results for the finite parts of individual $P$-term contributions, for $Z = 83$,
in units of $10^{-6}$.
\label{tab:fin:83}}
\begin{ruledtabular}
\begin{tabular}{lw{4.6}w{4.6}w{4.6}w{4.6}w{4.8}w{4.8}}
\multicolumn{1}{l}{Term} &
        \multicolumn{2}{c}{$\mathrm{IR'}$} &
            \multicolumn{2}{c}{$\mathrm{IR}$} &
                \multicolumn{1}{c}{R} &
                \multicolumn{1}{c}{Sum}
                \\[3pt]
\cline{2-3}
\cline{4-5}
\\[-7pt]
 & \multicolumn{1}{c}{$J \ge 0$}
 & \multicolumn{1}{c}{$J > 0$}
 & \multicolumn{1}{c}{$J \ge 0$}
 & \multicolumn{1}{c}{$J > 0$}
 &
 &
\\ \hline\\[-5pt]
$\mathrm{NW1}$  &  \cent{-}    &  \cent{-}    &  \cent{-} &  \cent{-} &    3.1250\,(5) &    3.1250\,(5)\\
$\mathrm{NW2}$  &  \cent{-}    &  \cent{-}    &  \cent{-} &  \cent{-} &   -3.8740\,(9) &   -3.8740\,(9)\\
$\mathrm{OW}$   &  \cent{-}    &  \cent{-}    &  \cent{-} &  \cent{-} &   -0.0502\,(1) &   -0.0502\,(1)\\
$\mathrm{ND1}$  & -19.5278     &  -1.7893     & -10.7545  & -1.1351   &  -11.1065\,(2) &  -14.0309\,(2)\\
$\mathrm{ND2}$  &  15.9184     &   1.4586     &  -8.9433  & -0.9439   &   -5.4600\,(10)&   -4.9453\,(10) \\ 
$\mathrm{ND3}$  &  \cent{-}    &  \cent{-}    &   5.3772  &  0.5676   &    5.9359\,(1) &    6.5035\,(1)\\
$\mathrm{NV1}$  &  17.1421     &  -0.5964     &  11.6501  & -0.4770   &    9.1974\,(6) &    8.1240\,(6)\\
$\mathrm{NV2}$  & -13.9737     &   0.4862     &   9.1883  & -0.3762   &    7.0464\,(11)&    7.1564\,(11)\\ 
$\mathrm{NV3}$  &  \cent{-}    &  \cent{-}    &  -2.0729  &  0.0849   &   -4.4407\,(1) &   -4.3558\,(1)\\
$\mathrm{OD}+\mathrm{OV}$  &  \cent{-} &  \cent{-}
                                              &   0.2691  &  0.2691   &    1.3896\,(21)&    1.6587\,(21) \\
$\mathrm{ADD}$  &  \cent{-}    &  \cent{-}    &  -7.5185  & -0.7936   &   -1.3635      &   -2.1571 \\
\hline\\[-5pt]
Sum             & -0.4410      &  -0.4410      &  -2.8044  & -2.8042   &    0.3995\,(28)&  -2.8459\,(28) \\
\end{tabular}
\end{ruledtabular}
\end{table*}

Final numerical results of our calculation of the $P$-term contribution
to the two-loop self-energy correction to the $1s$ $g$-factor
of H-like ions with $Z \ge 50$ are presented in Table~\ref{tab:total}.

\begin{table}
\caption{The $P$ term for the $1s$ $g$ factor of H-like ions,
in units of $10^{-6}$.
\label{tab:total}}
\begin{ruledtabular}
\begin{tabular}{lw{4.10}}
\multicolumn{1}{l}{Z} &
        \multicolumn{1}{c}{$E_P$}
\\
\hline\\[-5pt]
  50 &  -0.2674\,(42) \\
  60 &  -0.9498\,(83) \\
  70 &  -1.7257\,(31) \\
  83 &  -2.8459\,(28) \\
  92 &  -3.7369\,(115) \\
\end{tabular}
\end{ruledtabular}
\end{table}

In Table~\ref{tab:SESE} we summarize all contributions to the SESE
correction for the bound-electron $g$ factor of H-like tin ion ($Z = 50$), which was
recently measured \cite{morgner:23}.
Our final result of $-4.10\,(2) \times 10^{-6}$ is in good agreement with previous
estimations of this correction based on the $\Za$ expansion
\cite{pachucki:05:gfact,czarnecki:16,czarnecki:18,czarnecki:20,morgner:23},
$-4.3\,(3)\times 10^{-6}$, and improves its accuracy by an order of magnitude.

Our first complete results for the SESE correction in H-like tin were
reported in our recent Letter \cite{sikora:25}.
They improved the theoretical accuracy for the
$g$ factor of H-like tin by an order of magnitude over the
previous theoretical value reported in Ref.~\cite{morgner:23}.
The updated theoretical result showed a mild tension of
$2.1\sigma$ with the experimental value.
A potential cause for the tension could be an underestimated uncertainty
of the currently accepted nuclear charge radius of $^{118}$Sn, as listed in the tabulation
\cite{angeli:13}.
This interpretation is supported by the recent criticism by Ohayon
\cite{ohayon:25:radii}, who argued that the model dependence of the nuclear charge
distribution had not been properly accounted for in Ref.~\cite{angeli:13}
and their uncertainties should be substantially increased.

\section{Summary and outlook}

We presented a detailed description of our calculational approach developed for
the so-called $P$-term part of the two-loop self-energy correction for
the bound-electron $g$ factor. The $P$-term consists of Feynman diagrams
that need be treated in a mixed momentum-coordinate representation.
A detailed analysis of infrared divergencies present in individual $P$-term
contributions was carried out, and their cancelation was explicitly demonstrated.
Explicitly finite formulas were derived, suitable for numerical computations.
Numerical calculations of the $P$ term were carried out for the $g$ factor of the
$1s$ state of H-like ions with the nuclear charges $Z \ge 50$.
The results obtained for the $Z = 50$ allowed us to improve the
theoretical accuracy for the $g$ factor of H-like tin and enabled a stringent
test of the magnetic sector of bound-state QED theory.

In future studies, it will be necessary to extend our calculations of the SESE correction
to ions other than tin.
On the one hand, upcoming measurements on
$g$-factors of heavy H-like ions are planned by the ALPHATRAP experiment
\cite{sturm:19}, which require nonperturbative (in $\Za$) calculations of the SESE correction.
On the other hand, high-precision experimental results are already available for $g$ factors
of lower-$Z$ ions \cite{sturm:13:Si,sturm:14,sailer:22}, where the SESE correction
constitutes a dominant source of theoretical uncertainty.
An extension of the present calculations to the smaller nuclear
charges $Z < 50$ is likely to be rather challenging because
of the slower convergence of the partial-wave expansion and the stronger numerical
cancelations.
Possible strategies to overcome these difficulties could include
the partial-wave
convergence acceleration scheme \cite{sapirstein:23}, recently generalized to the two-loop
Lamb-shift calculations \cite{yerokhin:24:sese,yerokhin:25:sese},
as well as the use of the Coulomb gauge, which has proven
to be highly effective in one-loop self-energy calculations \cite{yerokhin:25:se}.

\section*{Acknowledgment}
This work is funded by the Deutsche Forschungsgemeinschaft
(DFG, German Research Foundation) under the Collaborative Research
Centre, Project-ID No. 273811115, SFB 1225 ISOQUANT.

\begin{table}
\caption{Two-loop self-energy contributions to the $1s$ bound-electron $g$-factor
of H-like tin, in units of $10^{-6}$. }
\label{tab:SESE}
\begin{ruledtabular}
\begin{tabular}{lw{4.10}c}
Term & \multicolumn{1}{c}{Value} & Ref. \\
\hline
F-term       & -4.0835\,(1)  & \cite{sikora:20} \\
LAL &           0.1086\,(35) & \cite{sikora:20,sikora:25} \\
M-term       &  0.143\,(17)    & \cite{sikora:25} \\
P-term       & -0.2674\,(42) & This work \\
Sum          & -4.099\,(19)  & This work, \cite{sikora:25} \\
$\Za$-expansion   & -4.25\,(30)   & \cite{pachucki:05:gfact,czarnecki:18,morgner:23}
\end{tabular}
\end{ruledtabular}
\end{table}


\appendix

\begin{widetext}

\section{Zeeman vertex function in momentum space}
\label{app:verzee}

The angular integrations in the Zeeman vertex function are performed in momentum space as explained
in Ref.~\cite{yerokhin:10:sehfs}. The result is expressed as
({\em cf.} Eq.~(30) of Ref.~\cite{yerokhin:10:sehfs})
\begin{align}
 \bra{n_1\kappa_1\mu_1} \gamma^0 \Lambda_{\rm Zee}(\vare_1,\vare_2) \ket{n_2\kappa_2\mu_2}
&\,
 =
   A_{\rho}\,\frac{\alpha}{48\pi^5}
 \int_0^{\infty} dp_1\,dp_2
   \int_{|p_1-p_2|}^{p_1+p_2}dq\,
   p_1p_2q\,e^{-q^2/\rho^2}\,(-1)^{(l_1-l_2)/2}
\nonumber \\
\times \Big\{
 &
  \Big[ -p_1\,K_1(\kappa_1,\mu_1;\kappa_2,\mu_2) + p_2\,K_1^{\prime}(\kappa_1,\mu_1;\kappa_2,\mu_2)\Big]
    {\cal R}_1(p_1,p_2,q)
\nonumber \\
 &
 + \Big[ -p_1\,K_1(-\kappa_1,\mu_1;-\kappa_2,\mu_2) + p_2\,K_1^{\prime}(-\kappa_1,\mu_1;-\kappa_2,\mu_2)\Big]
    {\cal R}_2(p_1,p_2,q)
\nonumber \\
 &
  -p_1p_2\,K_2(\kappa_1,\mu_1;\kappa_2,\mu_2)
    \Big[{\cal R}_3(p_1,p_2,q)+{\cal R}_4(p_1,p_2,q)\Big]
\nonumber \\
 &
  -p_1p_2\,K_2(-\kappa_1,\mu_1;-\kappa_2,\mu_2)
    \Big[{\cal R}_5(p_1,p_2,q)+{\cal R}_6(p_1,p_2,q)\Big]
    \Big\}\,,
\end{align}
where $\kappa_{1,2}$ and $\mu_{1,2}$ are the relativistic angular momentum quantum number and the
angular-momentum projection of the states $n_1$ and $n_2$, respectively,
${\cal R}_i(p_1,p_2,q)$ are the one-loop vertex functions given by
Eqs.~(A7)–(A12) of Ref.~\cite{yerokhin:99:sescr} and $K_1$, $K_1^{\prime}$, and $K_2$ are basic angular
integrals.

The angular integrals are defined as ({\em cf.} Eqs.~(A7a)-(A7c) of Ref.~\cite{yerokhin:10:sehfs})
\begin{align}
\frac{3i}{4\pi} \int d\hp_1\, d\hp_2\, & \ f(p_{1},p_{2},\xi)\,
     \chi^{\dag}_{\kappa_1 \mu_1}(\hp_1)\, [\hp_1\times \bsigma]_z\,
      \chi_{-\kappa_2\mu_2}(\hp_2)
= \int_{-1}^1 d\xi\,f(p_{1},p_{2},\xi)\,
        K_1(\kappa_1,\mu_1;\kappa_2,\mu_2)\,,
\\
\frac{3i}{4\pi} \int d\hp_1\, d\hp_2\, & \ f(p_{1},p_{2},\xi)\,
     \chi^{\dag}_{\kappa_1 \mu_1}(\hp_1)\, [\hp_2\times \bsigma]_z\,
      \chi_{-\kappa_2\mu_2}(\hp_2)
= \int_{-1}^1 d\xi\,f(p_{1},p_{2},\xi)\,
        K_1^{\prime}(\kappa_1,\mu_1;\kappa_2,\mu_2)\,,
\\
\frac{3i}{4\pi} \int d\hp_1\, d\hp_2\, & \ f(p_{1},p_{2},\xi)\,
     \chi^{\dag}_{\kappa_1 \mu_1}(\hp_1)\, [\hp_1\times \hp_2]_z\,
      \chi_{\kappa_2\mu_2}(\hp_2)
= \int_{-1}^1 d\xi\,f(p_{1},p_{2},\xi)\,
        K_2(\kappa_1,\mu_1;\kappa_2,\mu_2)\,,
\end{align}
where $\hp = \bfp/|\bfp|$, $\xi = \hp_1\cdot\hp_2$ and $f(p_{1},p_{2},\xi)$ is an arbitrary function.

The angular integrals $K_1$ and $K_1^{\prime}$ are evaluated as
\begin{align}\label{eq:K1}
        \left.
        \begin{array}{c}
        K_1(\kappa_1,\mu_1;\kappa_2,\mu_2) \\
        K_1^{\prime}(\kappa_1,\mu_1;\kappa_2,\mu_2)  \\
        \end{array}
        \right\}
=
 \sqrt{\frac32}\, (-1)^{j_1+\mu_1}\,
  C^{10}_{j_2\mu_2,j_1-\mu_1}\,
    S_{11}(\kappa_1,-\kappa_2)\,
    \times
       \left\{
        \begin{array}{c}
        P_{\overline{l}_2}(\xi) \\
        P_{{l}_1}(\xi) \\
        \end{array}
        \right.
      \,,
\end{align}
where $l_1 = |\kappa_1+1/2|-1/2$, $\overline{l}_2 = |\kappa_2-1/2|-1/2$, the angular coefficients $S_{JL}(\kappa_a,\kappa_b)$ are defined
by Eqs.~(C7)-(C9) of Ref.~\cite{yerokhin:99:sescr}, $C^{jm}_{j_1m_1,j_2m_2}$ are the Clebsch-Gordan
coefficients and $P_l(\xi)$ are the Legendre polynomials. For $\kappa_1 = \kappa_2$ and
$\mu_1 = \mu_2 = 1/2$, the above formula reproduces the results of Eq.~(A8a) of Ref.~\cite{yerokhin:99:sescr}.

The angular integral $K_2$ is calculated to be
\begin{align}\label{eq:K2}
        K_2(\kappa_1,\mu_1;\kappa_2,\mu_2) = &\,
        2\sqrt{2}\pi
            \sum_{L = L_{\rm min}}^{L_{\rm max}} P_L(\xi)\,
            \sum_{m = 0,\pm 1}\sum_{\sigma = \pm 1/2}
            C^{10}_{1m,1-m}\,
            C^{j_1\mu_1}_{l_1m_{1},1/2\sigma}\,
            C^{j_2\mu_2}_{l_2m_{2},1/2\sigma}\,
\nonumber \\ & \times
            R_3(l_1,m_{1},1,m,L,M)\,
            R_3(L,M,1,-m,l_2,m_{2})\,,
\end{align}
where $L_{\rm min} = \max(0,l_1-1,l_2-1)$, $L_{\rm max} = \min(l_1+1,l_2+1)$, and
\begin{align}
            R_3(l_1,m_1,l_2,m_2,l_3,m_3) = \int d\hp\, Y^*_{l_1m_1}(\hp)\,Y_{l_2m_2}(\hp)\,Y_{l_3m_3}(\hp)\,.
\end{align}
For $\kappa_1 = \kappa_2$ and
$\mu_1 = \mu_2 = 1/2$, the above formulas give the results of Eq.~(A8c) of Ref.~\cite{yerokhin:99:sescr}.
The dependence of $K_2$ on the projection of momenta is easily factorized as
\begin{align}
        K_2(\kappa_1,\mu_1;\kappa_2,\mu_2) =
         \frac{(-1)^{j_1-\mu_1}\,  C^{10}_{j_2\mu_2,j_1-\mu_1}}{(-1)^{j_1-\nicefrac{1}{2}}\,  C^{10}_{j_2\nicefrac{1}{2},j_1-\nicefrac{1}{2}}}
          \, K_2(\kappa_1,\nicefrac{1}{2};\kappa_2,\nicefrac{1}{2})\,,
\end{align}
so that it is sufficient to compute $K_2$ for $\mu_1 = \mu_2 = \nicefrac12$ only.

As a test, we calculate the matrix element of the regularized
magnetic potential $V_{g,\rho}$ in momentum space, which can be
expressed in a similar way,
\begin{align}
\bra{n_1\kappa_1\mu_1} V_{g,\rho} \ket{n_2\kappa_2\mu_1} =&\,
   A_{\rho}\,\frac{\alpha}{12\pi^4}
 \int_0^{\infty} dp_1\,dp_2
   \int_{|p_1-p_2|}^{p_1+p_2}dq\,
   p_1p_2q\,e^{-q^2/\rho^2}\,(-1)^{(l_1-l_2)/2}
\nonumber \\
\times \Big\{
 &
  \Big[ -p_1\,K_1(\kappa_1,\mu_1;\kappa_2,\mu_2) + p_2\,K_1^{\prime}(\kappa_1,\mu_1;\kappa_2,\mu_2)\Big]
    g_{n_1\kappa_1}(p_1)\, f_{n_2\kappa_2}(p_2)
\nonumber \\
 &
 + \Big[ -p_1\,K_1(-\kappa_1,\mu_1;-\kappa_2,\mu_2) + p_2\,K_1^{\prime}(-\kappa_1,\mu_1;-\kappa_2,\mu_2)\Big]
    f_{n_1\kappa_1}(p_1)\, g_{n_2\kappa_2}(p_2)
    \Big\}\,,
\end{align}
where $g_{n\kappa}(p)$ and $f_{n\kappa}(p)$ are the upper and the lower radial components of the Dirac wave function in momentum space,
see Eq.~(6) of Ref.~\cite{yerokhin:99:sescr}.

\section{Angular integration in the vertex overlapping contribution}
\label{app:POV}

For performing integrations over the angular variables in $E^P_{\mathrm{OV}}$, it is convenient to switch to
the spectral representation of the Green function. After angular integrations are carried out,
the resulting formulas can be easily rewritten in terms of radial components of the
Green function.
Furthermore, it is sufficient to consider the unsubtracted (``unsub'') contribution, since
the expressions for the subtraction term will be fully analogous.
We thus write Eq.~(\ref{eq:POV}), omitting the substraction part, as
\begin{align} \label{app2:1}
E^P_{\mathrm{OV}}(\mathrm{unsub}) = & \
    -4i\alpha \int_{C_F} d\omega
    \int \frac{d\bfp_1}{(2\pi)^3} \frac{d\bfp_2}{(2\pi)^3} \int d\bfz d\bfy\,
    \frac{e^{-i\bfq\cdot\bfz}}{\omega^2-\bfq^2 + i0}\,
    \sum_{n_1n_2}\frac1{(E-\vare_{n_1})(E-\vare_{n_2})}\,
    \psi_a^{\dag}(\bfz)\, \alpha_{\mu}\, \psi_{n_1}(\bfz)
\nonumber \\ & \times
    \psi^{\dag}_{n_1}(\bfy)\, V_g(\bfy)\,\psi_{n_2}(\bfy)\,
    \psi_{n_2}^{\dag}(\bfp_1)\,
         \gamma^0 \Gamma^{\mu}_R(E,\bfp_1;\vare_a,\bfp_2)\, \psi_a(\bfp_2)
         \,.
\end{align}
Following the procedure, described in Ref.~\cite{yerokhin:03:epjd} for the Lamb shift,
we transform the above formula as follows
\begin{align} \label{app2:2}
E^P_{\rm OV}&\,(\mathrm{unsub}) =
 -\frac{i\alpha^2}{2\pi^4} \int_{C_F} d\omega\,
 \int_0^{\infty} dp_1\,dp_2
   \int \frac{d\hp_1\,d\hp_2}{8\pi^2}\,
       \frac{(p_1p_2)^2}{\omega^2-\bfq^2+i0}
          \sum_{n_1n_2}\frac1{(E-\vare_{n_1})(E-\vare_{n_2})}
\nonumber \\ & \times
    \Bigg\{
    \sum_J i^{l_2-l_a-J}\,
      C_J(\kappa_1,\kappa_a)
          R^{(1)}_J(q,an_1)\,P(n_1n_2)\,
          \Big[
            {\cal F}^{n_2a}_1\,t_J(\kappa_1,\kappa_2,\kappa_a)
           + {\cal F}^{n_2a}_2\,t_J(\kappa_1,-\kappa_2,-\kappa_a)
 \Big]
\nonumber \\ &
 - \sum_{JL}
  i^{l_2-l_a-L+1} R^{(2)}_{JL}(q,an_1)\,P(n_1n_2)\,
       \Big[
       {\cal R}^{n_2a}_1\,s^{\sigma}_{JL}(\kappa_1,\kappa_2,-\kappa_a)
     +  {\cal R}^{n_2a}_2\,s^{\sigma}_{JL}(\kappa_1,-\kappa_2,\kappa_a)
\nonumber \\ &
     +p_1 {\cal R}^{n_2a}_3\,s^{p_1}_{JL}(\kappa_1,\kappa_2,\kappa_a)
     +p_2 {\cal R}^{n_2a}_4\,s^{p_2}_{JL}(\kappa_1,\kappa_2,\kappa_a)
     +p_1 {\cal R}^{n_2a}_5\,s^{p_1}_{JL}(\kappa_1,-\kappa_2,-\kappa_a)
     +p_2 {\cal R}^{n_2a}_6\,s^{p_2}_{JL}(\kappa_1,-\kappa_2,-\kappa_a)
     \Big]
     \Bigg\}\,,
\end{align}
where the functions $ {\cal F}^{n_2a}_i$ and ${\cal R}^{n_2a}_i$ come from
the vertex operator sandwiched between the Dirac wave functions
(see Eqs.~(147) and (148) of Ref.~\cite{yerokhin:03:epjd}) and
the angular coefficients are defined by (cf. Eqs.~(150) and (151) of Ref.~\cite{yerokhin:03:epjd})
\begin{align}
t_J(\kappa_1,\kappa_2,\kappa_a) =
 \sum_{\mu_1\mu_2 M}
      (-1)^{j_1-\mu_1}\,C^{10}_{j_2\mu_2,j_1-\mu_1}\,
    s_{JM}^{n_1a}\,
    \chi^{\dag}_{\kappa_2 \mu_2}(\hp_1)\, Y_{JM}(\hq)\,
        \chi_{\kappa_a \mu_a}(\hp_2) \,,
\\
s_{JL}^{\sigma}(\kappa_1,\kappa_2,\kappa_a) =
 \sum_{\mu_1\mu_2 M}
      (-1)^{j_1-\mu_1}\,C^{10}_{j_2\mu_2,j_1-\mu_1}\,
    s_{JM}^{n_1a}\,
    \chi^{\dag}_{\kappa_2 \mu_2}(\hp_1)\, {\vec \sigma}\cdot \vec{Y}_{JLM}(\hq)\,
        \chi_{\kappa_a \mu_a}(\hp_2) \,,
\\
s_{JL}^{p_i}(\kappa_1,\kappa_2,\kappa_a) =
 \sum_{\mu_1\mu_2 M}
      (-1)^{j_1-\mu_1}\,C^{10}_{j_2\mu_2,j_1-\mu_1}\,
    s_{JM}^{n_1a}\,
    \chi^{\dag}_{\kappa_2 \mu_2}(\hp_1)\, {\hp}_i\cdot \vec{Y}_{JLM}(\hq)\,
        \chi_{\kappa_a \mu_a}(\hp_2) \,.
\end{align}
Furthermore,
the radial integrals are (cf. Eqs.~(145) and (146) of  Ref.~\cite{yerokhin:03:epjd})
\begin{align}
R^{(1)}_J(q,an) =&\, C_J(\kappa_n,\kappa_a) \int_0^{\infty} dx\, x^2 j_J(qx)
         (g_ag_n+f_af_n)\,,
    \\
R^{(2)}_{JL}(q,an) =&\,  \int_0^{\infty} dx\, x^2 j_L(q x)\,
 \Bigl[g_af_n
 S_{JL}(\kappa_a,-\kappa_n)
    -f_ag_n S_{JL}(-\kappa_a,\kappa_n)\Bigr]\,,
    \\
P(n_1,n_2) =&\, 2\,\frac{-\kappa_1-\kappa_2}{\sqrt{3}}\,C_1(-\kappa_2,\kappa_1)
 \,\int_0^{\infty} dx\, x^3
         (g_{n_1}f_{n_2}+f_{n_1}g_{n_2})\,,
\end{align}
where $q = |\bfq|$, $g_a=g_a(x)$ and $f_a = f_a(x)$ are components of
radial wave functions, and $j_l(z)$ is the spherical Bessel function.
The standard angular coefficients $C_{L}(\kappa_1,\kappa_2)$
and $S_{JL}(\kappa_1,\kappa_2)$ can be found, e.g., in Appendix C of
Ref.~\cite{yerokhin:03:epjd}.

Now we turn to the integration
over $\hp_1$ and $\hp_2$ in Eq.~(\ref{app2:2}). It is complicated by the fact
that we have to integrate out all angles except for $\xi = \hp_1\cdot \hp_2$
analytically, whereas the remaining integration over $\xi$ can be performed
only numerically.
We perform angular integration with the help of the following identity
(see Appendix C of Ref.~\cite{yerokhin:20:gfact})
\begin{align}\label{app2:3}
\frac{1}{8\pi^2} \int d\hp_1d\hp_2\, f(q)\, G(\hp_1,\hp_2) =
 \int_{-1}^1d\xi\, f(q)\, \hat{G}(\xi)\,,
\end{align}
where $f(q)$  and $G(\hp_1,\hp_2)$ are arbitrary functions of the
specified arguments and
\begin{align}\label{app2:4}
\hat{G}(\xi) = \frac1{4\pi} \sum_{lm}P_l(\xi)
 \int d\hp_1d\hp_2\, Y_{lm}(\hp_1)\,Y^*_{lm}(\hp_2)\,G(\hp_1,\hp_2)\,.
\end{align}
With help of Eqs.~(\ref{app2:3}) and (\ref{app2:4}) we can integrate
out all angles except $\xi$ in the angular coefficients $t_J$, $s^{\sigma}_{JL}$,
$s^{p_{1,2}}_{JL}$, in a similar way to that described in Eqs.~(C14)-(C20) of
Ref.~\cite{yerokhin:20:gfact}. The calculation is quite straightforward but
tedious; the resulting formulas are too long to be presented here explicitly.
What is unfortunate, however, is not the length of the formulas, but their numerical
instability in the region $q\to 0$ and for large $J$. Because of
this we had to employ the extended-precision arithmetic (up to 64 decimal digits) for computation
of these angular coefficients. Even so, we were able to extend the partial-wave
expansion in $E^P_{\rm OV}$ only up to $|\kappa_{\rm max}| = 15$, since for
higher partial waves even the 64-digit arithmetic was not sufficient.
This situation contrasts with the case of the Lamb shift, where a numerically stable
algorithm for computing the analogous angular factors was found, 
as described in Appendix C of Ref.~\cite{yerokhin:10:sese}.
Unfortunately, we have not been able to extend this stable algorithm to the
$g$-factor case.

\end{widetext}

\end{document}